\documentclass[prl,twocolumn,superscriptaddress]{revtex4-1} 
\usepackage[latin1]{inputenc}
\usepackage{graphicx}
\usepackage{amssymb}
\usepackage{amsmath}
\usepackage{xspace}
\usepackage{dcolumn}
\usepackage{bm}
\usepackage{color}
\usepackage[extra]{tipa}
\usepackage{cancel}

\newfont{\tensy}{cmsy10}

\newcommand{\ie}[0]{i.e.\@\xspace}

\newcommand{\rmi}{\text{i}}
\newcommand{\UP}[0]{\uparrow}
\newcommand{\DO}[0]{\downarrow}

\newcommand{\on}{\hat{n}}
\newcommand{\oP}{\hat{P}}
\newcommand{\oQ}{\hat{Q}}

\newcommand{\bk}{\boldsymbol{k}}

\newcommand{\kB}{k_\text{B}}
\newcommand{\nag}{{\phantom{\dag}}}

\newcommand{\las}[0]{\langle}
\newcommand{\ras}[0]{\rangle}

\begin{document}

\author{Chuang Chen}
\affiliation{Beijing National Laboratory for Condensed Matter Physics and Institute of Physics, Chinese Academy of Sciences, Beijing 100190, China}
\affiliation{School of Physical Sciences, University of Chinese Academy of Sciences, Beijing 100190, China}
\author{Xiao Yan Xu}
\affiliation{Department of Physics, Hong Kong University of Science and Technology, Clear Water Bay, Hong Kong, China}
\author{Zi Yang Meng}
\affiliation{Beijing National Laboratory for Condensed Matter Physics and Institute of Physics, Chinese Academy of Sciences, Beijing 100190, China}
\affiliation{CAS Center of Excellence in Topological Quantum Computation and School of Physical Sciences,
University of Chinese Academy of Sciences, Beijing 100190, China}
\affiliation{Songshan Lake Materials Laboratory, Dongguan, Guangdong 523808, China}
\author{Martin Hohenadler}
\affiliation{\mbox{Institut f\"ur Theoretische Physik und Astrophysik,
    Universit\"at W\"urzburg, 97074 W\"urzburg, Germany}}


\title{Charge-Density-Wave Transitions of Dirac Fermions Coupled to Phonons}

\begin{abstract}
The spontaneous generation of charge-density-wave order in a Dirac fermion
system via the natural mechanism of electron-phonon coupling is studied in
the framework of the Holstein model on the honeycomb lattice. Using 
two independent and unbiased quantum Monte Carlo methods, the phase diagram
as a function of temperature and coupling strength is determined. It features a
quantum critical point as well as a line of thermal critical
points. Finite-size scaling appears consistent with fermionic Gross-Neveu-Ising
universality for the quantum phase transition, and bosonic Ising universality
for the thermal phase transition. The critical temperature 
has a maximum at intermediate couplings. Our findings motivate experimental efforts to
identify or engineer Dirac systems with sufficiently strong and tunable
electron-phonon coupling.
\end{abstract} 

\date{\today}

\maketitle

The experimental advances in preparing single-layer graphene
\cite{Neto_rev} 
have put Dirac fermions at the focus of condensed matter physics. While the
single-electron properties are relatively well understood, correlation
effects remain a highly active area of research \cite{tang2018role}. Due to
the two-dimensional (2D) nature of the problem,
theoretical models can be analyzed by powerful theoretical and numerical
methods, offering the prospect of a comprehensive understanding. The field
has recently received another boost by the remarkable properties of 
other honeycomb systems, in particular quantum-spin-Hall physics
in bismuthene \cite{reis2017bismuthene} and unconventional superconductivity
in twisted bilayer graphene \cite{cao2018unconventional}. Finally, 
massive Dirac phases such as charge-density-wave (CDW) insulators in transition-metal
dichalcogenides \cite{manzeli20172d} promise future applications in
optoelectronics.

Theoretical studies of massive $(2+1)$D Dirac fermions
were pioneered by Semenoff \cite{PhysRevLett.53.2449}, who considered a staggered
fermion density or CDW, and Haldane \cite{Haldane98}, who introduced
a topological mass that produces an integer quantum Hall state 
in the absence of a magnetic field. Such problems become even richer if the
masses arise from spontaneous symmetry breaking at interaction-driven phase transitions.
Particularly remarkable aspects of Dirac systems are that (i)
phase transitions occur at nonzero critical values and
(ii) the gapless fermionic excitations can strongly modify the critical
behavior, giving rise to fermionic quantum critical points 
\cite{PhysRevLett.97.146401,Herbut09a,scherer2016gauge,zhou2016mott,li2017fermion,
classen2017fluctuation,he2018dyn,torres2018fermion,xu2018kekule,lang2018quantum}. The
interplay of different order parameters provides a route to 
deconfined quantum critical points \cite{senthil2004decon} 
and emergent symmetries \cite{sato2017dirac,qin2017dual} (see
Ref.~\cite{wang2017decon} for a review).

Numerous interactions have been explored numerically in the framework of 
honeycomb lattice models. 
A sufficiently strong onsite Hubbard repulsion 
yields an antiferromagnetic Mott insulator
\cite{Sorella92,Sorella12,Assaad13}. The same holds for a
more realistic $1/r$ Coulomb repulsion, although
the nonlocal part of the interaction---relevant for graphene where screening is
absent---enhances CDW fluctuations \cite{PhysRevB.90.085146}.
A dominant nearest-neighbor repulsion favors a CDW state 
\cite{Raghu08,PhysRevB.89.111101,1367-2630-16-10-103008,li2015fermion,Wessel2016}
but is rather unrealistic; for spinful fermions, quantum Monte
Carlo (QMC) simulations are hampered by the sign problem.
Mean-field predictions of interaction-generated topological states in
extended Hubbard models \cite{Raghu08} inspired significant efforts
to address fluctuation effects. For spinless fermions, unbiased numerical
methods reveal the absence of topological phases but support CDW, valence
bond solid, and charge-modulated ground states (see
Ref.~\cite{capponi2016phase} for a review). Similar conclusions were recently
reached for the spinful problem \cite{PhysRevB.95.085143,PhysRevB.97.125142}.
Finally, bond-bond interactions were found to produce valence bond,
antiferromagnetic, quantum-spin-Hall, and CDW states
\cite{chandrasekharan2013,li2017fermion,he2018dyn}.

Here, we consider electron-phonon coupling as the mechanism for
CDW order. QMC investigations along these lines have so far been restricted by the challenges
in simulating electron-phonon models, as addressed by several 
recent methodological advances
\cite{PhysRevB.98.085405,PhysRevB.98.041102,BaSc2018,Ka.Se.So.18}. We carried
out large-scale QMC simulations of the fundamental Holstein molecular-crystal
model \cite{Ho59a} to determine the phase diagram as a function
of coupling strength and temperature. Moreover, we investigate the nature of the
observed quantum and thermal phase transitions.

{\it Model.}---Within the Holstein model, electrons coupled to quantum phonons
on the honeycomb lattice are described by the Hamiltonian 
\begin{align}\label{eq:model}
  \hat{H}
  &=
    -t \sum_{\las ij \ras\sigma} \hat{c}^{\dagger}_{i\sigma} \hat{c}^\nag_{j\sigma}
    +
    \sum_{i}
    \left[
    \frac{1}{2M}
    \oP^2_{i}
    +
    \frac{\kappa}{2}\oQ_{i}^2
    \right]
    -
    g
    \sum_{i} \hat{Q}_{i} 
     \hat{\rho}_i
    \,.
\end{align}
The first term represents nearest-neighbor electronic hopping, the second
term independent Einstein phonons at each lattice site, and the third term the
coupling between fluctuations of the local electron number $\hat{\rho}_i=\on_i-1$
and the lattice displacement $\oQ_i$. Here, $\on_{i} = \sum_\sigma
\hat{c}^\dag_{i\sigma}\hat{c}^\nag_{i\sigma}$, the phonon frequency
$\omega_0=\sqrt{\kappa/M}$, and we introduce the dimensionless coupling
$\lambda=g^2/(\kappa W)$ with the free bandwidth $W=6t$. We consider
half-filling and work in units where $\kB$, $\hbar$ and the lattice constant
are equal to one.

For $\lambda=0$, Eq.~(\ref{eq:model}) gives the well-known semimetallic
band structure $\epsilon(\bm{k})$ with linear excitations at the Dirac points
$K,K'$ \cite{Neto_rev}. An expansion around these points  yields
a Dirac equation in terms of eight-component spinor fields corresponding to $N=2$
(spin $\UP,\DO$) Dirac fermions with two flavors (valleys $K,K'$) and two pseudospin
directions (sublattices A,B) \cite{Neto_rev}. 

{\it Methods.}---We used the determinant QMC (DQMC) \cite{Blankenbecler81} and the continuous-time interaction expansion
(CT-INT) QMC methods \cite{Rubtsov05}. In the former, the electrons are
integrated out and the phonons are sampled using local and block updates \cite{scalettar1991,johnston2013} as
well as global moves based on an effective bosonic model determined by a
self-learning
scheme~\cite{PhysRevB.95.041101,PhysRevB.95.241104,Xu2017SLMC,Nagaiself2017,PhysRevB.98.041102},
see SM \cite{SM}.  In 
CT-INT, the phonons are integrated out and
the resulting electronic model with a retarded interaction is sampled
\cite{Assaad07}. While CT-INT works in continuous imaginary time, 
a Trotter discretization $\Delta\tau=0.1$ was used for DQMC calculations. Although both
methods are in principle capable of simulating any parameters, CT-INT is most
efficient at weak coupling and less problematic with respect to
autocorrelations \cite{PhysRevB.98.085405}. DQMC simulations require more care regarding
the sampling but---especially in  combination with self-learning---can
access stronger couplings and larger system sizes. We used lattices with
$L\times L$ unit cells ($2L^2$
sites) and $L=3n$ ($n=1,2,\dots$) whose reciprocal lattice contains the Dirac
points that determine the low-energy physics. 

\begin{figure}
  \includegraphics[width=0.45\textwidth]{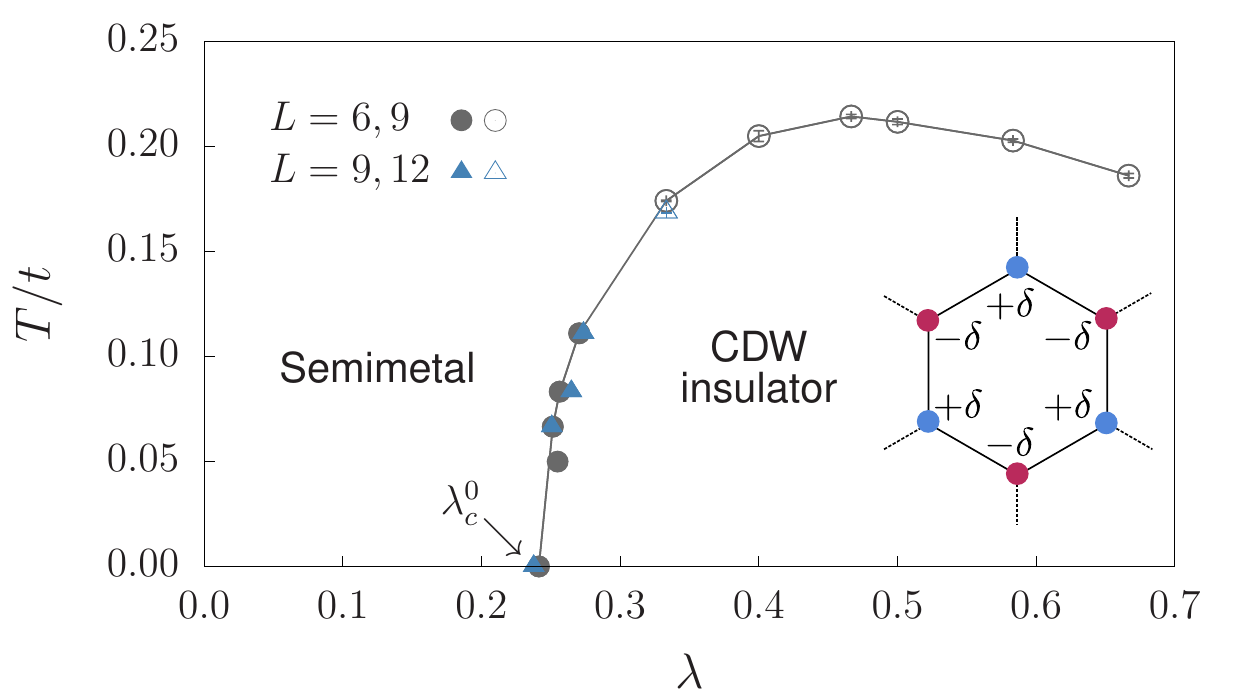}
  \caption{\label{fig:phasediagram} Phase diagram of the Holstein model~(\ref{eq:model}) 
    for $\omega_0=0.5t$. CDW order with a staggered 
    charge disproportionation $\pm\delta$ (inset) exists beyond 
    a quantum critical point at $\lambda_c^0\approx0.2375$ and below
    a critical temperature $T_c(\lambda)$. 
    Critical values were obtained from the crossings of the correlation ratio
$R_\text{c}$ for different system sizes $L$ as a function of $\lambda$
(filled symbols) or $T$ (open symbols), respectively. Data obtained from
CT-INT ($T\leq 0.05t$) and DQMC ($T>0.05t$) simulations, respectively.
The line is a guide to the eye.
 }
\end{figure}

{\it Phase diagram.}---The existence of CDW order at sufficiently strong
coupling can be inferred from two opposite limits. For classical phonons
($\omega_0=0$), we can make a mean-field ansatz $\oQ_i\mapsto
(-1)^{i}\overline{Q}$, corresponding to a staggered chemical potential or
Semenoff mass that breaks the sublattice and chiral symmetry
\cite{PhysRevLett.53.2449}. The lattice
displacements are accompanied by a density imbalance $\delta=|\las
\on^\text{A}\ras -\las \on^\text{B}\ras|$ (see inset of Fig.~\ref{fig:phasediagram}). The
band structure acquires a
gap at the Fermi level, $E(\bm{k}) = \pm\sqrt{\epsilon^2(\bm{k})+\Delta^2}$.
Spontaneous mass generation is described by a gap
equation identical to that for the Mott transition of the Hubbard model upon
identifying $\overline{Q}=m/2$ ($\Delta=g\overline{Q}$), $\lambda W=U$.
The mean-field critical value is $U_c=2.23t$ or $\lambda_c=0.37$
\cite{Sorella92}, which may be
compared to $U_c\approx 3.8t$ or $\lambda_c\approx0.63$ from QMC simulations
\cite{Assaad13,Sorella12,Toldin14}. The nonzero
critical value reflects the stability of the semimetal at weak coupling
\cite{PhysRevLett.97.146401}, the
origin of which is the linearly vanishing density of states,
$N(\omega)\sim|\omega|$ \cite{Neto_rev}.

In the opposite, antiadiabatic limit $\omega_0\to\infty$, integrating out the
phonons in the path-integral representation yields an attractive Hubbard
model with $U=\lambda W$ \cite{Hirsch83a}. By symmetry \cite{yang1990so},
$U_c$ has the same magnitude as for the Mott transition of the repulsive
Hubbard model, namely $3.8t$ \cite{Assaad13,Sorella12,Toldin14}. Under the
Lieb-Mattis particle-hole transformation that yields $U\to-U$, the order
parameters for CDW and superconductivity of the attractive Hubbard model
combine into a 3D vector that maps to the magnetization of the
repulsive model \cite{yang1990so}. This implies (i) coexistence of CDW order and
superconductivity for $U>U_c$ \cite{PhysRevLett.97.230404} and (ii) long-range order that spontaneously
breaks the SO(3) symmetry only at $T=0$ \cite{Hirsch85}.
An expansion in $1/\omega_0$ in the path-integral representation of the
Holstein model produces terms that violate the SO(3) symmetry \cite{Hirsch83a}.
A mean-field decoupling with an Ising CDW order parameter---reflecting the
two possible choices for the sign of the excess charge $\delta$ in
Fig.~\ref{fig:phasediagram}---gives again $U_c=2.23t$ or $\lambda_c=0.37$.
However, while Ising-like CDW order in the square-lattice Holstein model is strongly
suggested by the nesting-related, stronger divergence of the CDW
susceptibility compared to pairing \cite{PhysRevB.42.2416}, we are not aware
of such an argument for the honeycomb Holstein model considered here.

For quantitative insights into the experimentally relevant case of finite
$\omega_0$, we turn to QMC simulations. We focus on
$\omega_0=0.5t$, for which both quantum fluctuations and retardation effects
are significant. We determined critical values either at fixed coupling or at
fixed temperature. The values reported in Fig.~\ref{fig:phasediagram} are
based on the renormalization-group invariant correlation ratio
$R_\text{c}=1-S_\text{c}(\bm{Q}+\delta\bm{q})/S_\text{c}(\bm{Q})$ \cite{Binder1981}
calculated from the charge structure factor
$S_\text{c}(\bm{q})=L^{-2}\sum_{ij} e^{-\rmi
  \bm{q}\cdot(\bm{r}_i-\bm{r}_j)}\las (\on^\text{A}_i-\on^\text{B}_i)(\on^\text{A}_j-\on^\text{B}_j) \ras$. The CDW
order is within the unit cell, so the ordering wavevector
$\bm{Q}=\mathit{\Gamma}=(0,0)$. If $\bm{Q}+\delta\bm{q}$ is a neighboring
point in the Brillouin zone, long-range order and hence a divergence of
$S_\text{c}(\mathit{\Gamma})$ implies $R_\text{c}\to 1$ for
$L\to\infty$, otherwise $R_\text{c}\to 0$. At the critical point,
$R_\text{c}$ is independent of $L$ up to scaling corrections, so that the
critical value can be estimated from intersections of $R_\text{c}$ for
different $L$. Crucially, the scaling holds independent of any 
critical exponents and $R_\text{c}$ usually has smaller scaling corrections
than $S_\text{c}(\mathit{\Gamma})$ \cite{Binder1981,PhysRevLett.117.086404}.

\begin{figure}
  \includegraphics[width=0.225\textwidth]{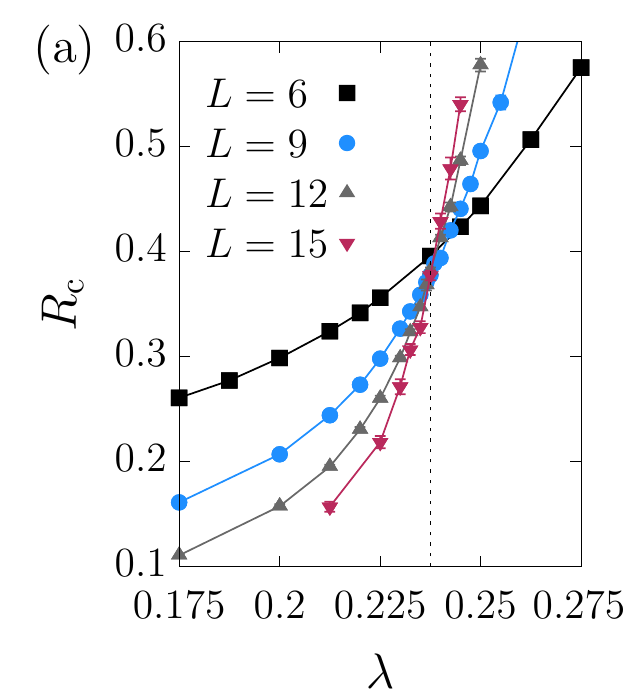}
  \includegraphics[width=0.225\textwidth]{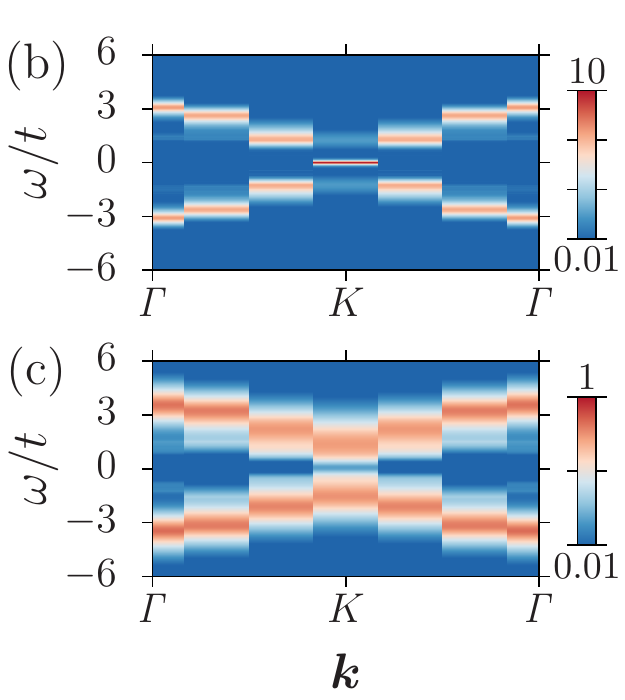}
  \caption{\label{fig:criticalvalues} (a) Estimation of the critical value
    $\lambda^0_c\approx0.2375$ for the quantum critical point from the
    intersections of the correlation ratio $R_\text{c}$. Here, $\beta t=L$,
    $\omega_0=0.5t$. Single-particle spectral function $A(\bk,\omega)$ in (b) the
    semimetallic phase ($\lambda=0.1$) and (c) the CDW
    phase ($\lambda=0.4$) for $\beta t=L=9$. Results were obtained with the
    CT-INT method.}
\end{figure}

Near the quantum critical point, the RG-invariant correlation ratio $R_\text{c}$ depends
on $(\lambda-\lambda_c)L^{1/\nu}$ and $L^z/\beta$. For the finite-size
scaling analysis, we took $\beta t=L$ (\ie, $z=1$) based on the expected emergent Lorentz
symmetry \cite{Herbut09}.
Figure~\ref{fig:criticalvalues}(a) suggests a critical value
$\lambda^0_c\approx0.2375$. Similar analysis for other parameters yields the
phase boundary in Fig.~\ref{fig:phasediagram}, shown in terms of
the intersections of $L=6,9$ and $L=9,12$, respectively.
Apart from the absence of long-range order at $\lambda<\lambda^0_c$
[Fig.~\ref{fig:criticalvalues}(a)], the CDW transition is also apparent in
the single-particle spectral function $A(\bk,\omega)$ \cite{SM}. We find
gapless excitations at the Dirac point for $\lambda=0.1$
[Fig.~\ref{fig:criticalvalues}(b)] and a gap
at the Fermi level for $\lambda=0.4$ [Fig.~\ref{fig:criticalvalues}(c)].
We found no evidence of long-range superconducting order for the
  parameters considered \cite{SM}.
 
In Fig.~\ref{fig:phasediagram}, CDW order persists up to a critical
temperature $T_c$. After an initial increase, asymptotically 
determined by the quantum critical point via $T_c \sim
|\lambda-\lambda^0_c|^{z\nu}$ \cite{Wessel2016},
$T_c$ takes on a maximum
before decreasing at even stronger couplings \cite{PhysRevB.63.115114}. This
can be understood within an effective $t$-$V$ model of singlet bipolarons
(hardcore bosons) \cite{Hirsch83a}. The binding energy of the latter 
continues to grow with $\lambda$, but their exchange interaction $V$
that sets the temperature for CDW order in this regime decreases
(cf. $T_c\sim J$ for the Ising model).
An expression for $V$ in the Holstein model is given in
Ref.~\cite{Hirsch83a} and simplifies to $V\sim t/\lambda$ for $\omega_0\gg
t$. The observed decrease of $T_c$ with increasing electron-phonon
coupling $\lambda$ contrasts the linear increase of $T_c$ with increasing
electron-electron repulsion in models for CDW order from Coulomb repulsion
\cite{PhysRevB.32.103,Wessel2016,PhysRevB.95.035108}.
Finally, the phase boundary is expected to shift to stronger couplings at larger
$\omega_0$ due to enhanced lattice fluctuations, reaching
$\lambda^0_c\approx0.63$ \cite{Assaad13,Sorella12,Toldin14} 
in the Hubbard limit $\omega_0\to\infty$ where $T_c\equiv0$ for any
$\lambda>\lambda_c^0$ due to the continuous SO(3) symmetry.

{\it Quantum phase transition.}---In Dirac systems, the Yukawa coupling
between the gapless fermions and order parameter fluctuations
described by Gross-Neveu field theories gives rise to fermionic critical
points rather than Wilson-Fisher bosonic
critical points \cite{PhysRevLett.97.146401,Herbut09a}. Gross-Neveu-Ising
universality for CDW transitions was previously observed for
$N=1$ Dirac fermions with nearest-neighbor
Coulomb repulsion \cite{PhysRevB.89.111101,1367-2630-16-10-103008,li2015fermion,Wessel2016},
and $N=2$ Dirac fermions with bond interactions \cite{chandrasekharan2013,he2018dyn}.
For the Holstein model, Gross-Neveu-{\it Heisenberg} universality is well established
\cite{Assaad13,Toldin14,Otsuka16} for $\omega_0\to\infty$, where it maps to
the attractive Hubbard model. The $3+1$ dimensional Gross-Neveu theory for the adiabatic
limit $\omega_0\to0$ should have a correlation length exponent $\nu=1/2$
\cite{Assaad13}. For general $\omega_0$, $2+1$ dimensional, $N=2$ Gross-Neveu-Ising universality
is expected.

\begin{figure}
  \includegraphics[width=0.45\textwidth]{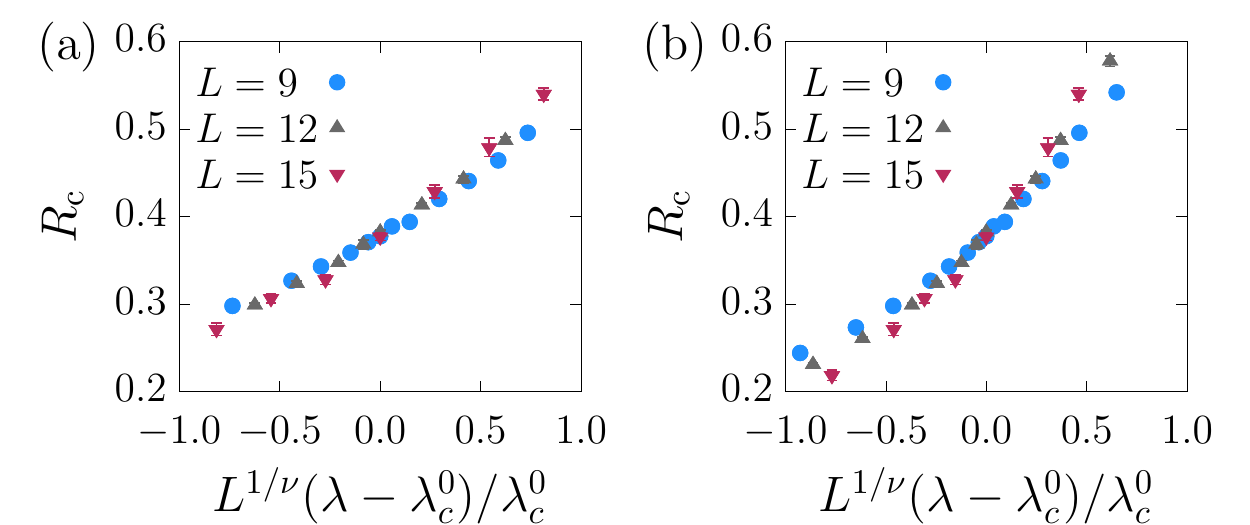}
  \caption{\label{fig:QPT} 
   Scaling collapse of the correlation ratio $R_\text{c}$ using
   $\lambda^0_c=0.2375$ and (a) $1/\nu=1.2$, (b) $1/\nu=0.931$.   
}
\end{figure}

For a preliminary analysis, we use $\lambda^0_c=0.2375$ from
Fig.~\ref{fig:criticalvalues}(a) and available estimates for the exponent
$\nu$ from QMC simulations ($1/\nu=1.2(1)$ \cite{chandrasekharan2013})
and the $\epsilon$-expansion ($1/\nu=0.931$
\cite{PhysRevD.96.096010}), respectively. The rescaled correlation ratio for
$L=9,12,15$ in Fig.~\ref{fig:QPT} appears more consistent with $1/\nu=1.2$
[Fig.~\ref{fig:QPT}(a)] than with $1/\nu=0.931$ [Fig.~\ref{fig:QPT}(b)].
As a further consistency check, we determined $\lambda^0_c$ from the best scaling collapse
\cite{autoscale} on the interval $[-1,1]$. The exponent $1/\nu=0.931$ yields
$\lambda^0_c\approx 0.239(2)$, whereas $1/\nu=1.2$ yields $\lambda^0_c\approx
0.238(1)$, slightly closer to the value obtained in Fig.~\ref{fig:criticalvalues}(a)
without any assumption about the value of $\nu$.

A direct estimate of $\nu$ based on an improved data set appears feasible and
is motivated by the rather different existing results \cite{PhysRevD.96.096010}.  
At the same time, a potential additional
complication---absent in purely fermionic models ---is that the phonon frequency
interpolates between three different fixed points, namely 
mean-field scaling ($\nu=1/2$ \cite{Assaad13})
at $\omega_0=0$, Gross-Neveu-Ising scaling for $\omega_0>0$, and
Gross-Neveu-Heisenberg scaling for $\omega_0=\infty$. For $\omega_0=0.5t$,
the proximity to the adiabatic fixed point may give rise to
crossover effects in the exponents. Another interesting possibility that has
to be ruled out is the formation of singlet pairs---triggered by the
attractive component of the frequency-dependent fermion-fermion
interaction---prior to the CDW transition, as in the 1D Holstein model
\cite{PhysRevB87.075149}. In the absence of gapless fermion
excitations at $\lambda_c$, Wilson-Fisher theory suggests $2+1=3$ dimensional
Ising universality. 
Both the expected $\omega_0=0$ value ($1/\nu=2$ \cite{Assaad13}) and the 3D
Ising value ($1/\nu\approx1.59$ 
\cite{PhysRevB.59.11471}) are larger than predicted for
the $N=2$ Gross-Neveu-Ising universality class \cite{PhysRevD.96.096010}.

\begin{figure}[t]
  \includegraphics[width=0.45\textwidth]{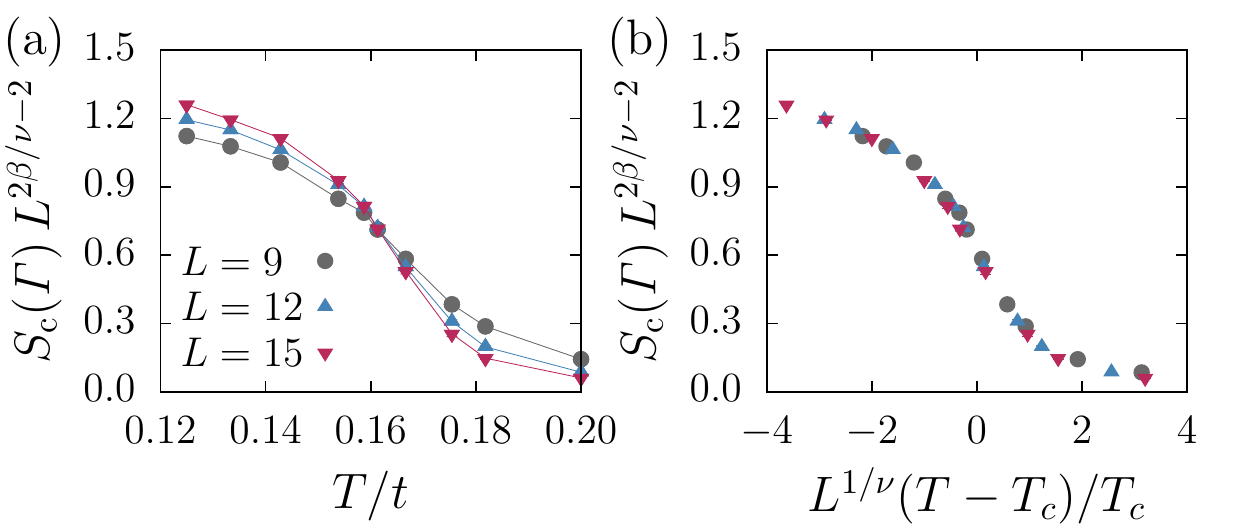} 
  \caption{\label{fig:2DIsing} Finite-size scaling of the density structure
    factor using the 2D Ising critical exponents $\beta=1/8$, $\nu=1$.
    The critical temperature $T_c=0.162t$ obtained from the best scaling
    collapse shown in (b) is consistent with the crossing point in (a).
    Here, $\omega_0=0.5t$, $\lambda=1/3$. Results obtained with the
    self-learning DQMC method.
}
\end{figure}

{\it Thermal phase transition.}---Starting from the CDW ground state
at $\lambda>\lambda_c^0$, long-range order is destroyed by thermal fluctuations
at $T_c$. The phase transition is expected to exhibit 2D Ising
universality with critical exponents $\beta=1/8$ and
$\nu=1$. Figure~\ref{fig:2DIsing}(a) shows that for $\omega_0=0.5t$ and $\lambda=1/3$,
the rescaled charge structure factor has a crossing of different system sizes
compatible with  $T_c=0.159(2)$ in Fig.~\ref{fig:phasediagram}. The best scaling
collapse on the interval $[-2,2]$ produces $T_c=0.1648(5)t$ and is shown in
Fig.~\ref{fig:2DIsing}(b).

{\it Discussion.}---Our investigation of spontaneously generated CDW order
from electron-phonon coupling on the honeycomb lattice  reveals several
differences to previous work. Perhaps most importantly, the Dirac
band structure gives rise to a quantum critical point with expected
Gross-Neveu-Ising universality at nonzero
coupling. In contrast, the Fermi liquid of the square lattice is expected to have a
weak-coupling instability due to perfect nesting and a van Hove singularity
\cite{PhysRevB.42.2416,PhysRevB.98.085405}. The thermal CDW transition
appears to have the same Ising universality as for the square lattice
\cite{PhysRevB.98.085405,PhysRevLett.120.187003,PhysRevB.98.041102,BaSc2018}. Such a
transition is absent in
the antiadiabatic limit, corresponding to the attractive Hubbard
model. While the latter is useful to describe superconductivity away from
half-filling, it supports long-range CDW order only at $T=0$ \cite{Hirsch85}. 
Models with dominant nearest-neighbor repulsion capture the finite-temperature
CDW transition \cite{PhysRevB.32.103,Wessel2016,PhysRevB.95.035108} but 
not the suppression of $T_c$ at strong coupling. Finally, we showed
that, similar to the square lattice, CDW order prevails over
superconductivity at half-filling.

{\it Outlook.}---There are several interesting future directions. The
fermionic quantum criticality requires additional efforts.
Superconductivity at nonzero doping and the
competition between CDW order and antiferromagnetism in a Holstein-Hubbard
model should be investigated.
Our work may also provide a starting point for more realistic modeling of 
twisted bilayer graphene \cite{cao2018unconventional} or transition-metal
dichalcogenides \cite{manzeli20172d}. On the experimental side, a key
question is if CDW order from electron-phonon coupling can be realized in one of
the many Dirac systems currently being investigated.

While writing this Letter, we became aware of a closely related study of the
same model whose results are fully consistent with ours \cite{PhysRevLett.122.077602}.

\begin{acknowledgments}
We thank F. Assaad and I. Herbut for helpful discussions.
MH acknowledges support by the DFG through SFB 1170 ToCoTronics; CC and ZYM
by the Ministry of Science and Technology of China
through the National Key Research and Development Program
(grant~2016YFA0300502), the Strategic Priority Research Program of the Chinese
Academy of Sciences (XDB28000000), and the National Science
Foundation of China (11574359); XYX by HKRGC (C6026-16W,
16324216, 16307117). We thank the John von Neumann
Institute for Computing (NIC) for computer resources on 
JURECA~\cite{Juelich} at the J\"ulich Supercomputing Centre (JSC),
the Center for Quantum Simulation Sciences at Institute of Physics,
Chinese Academy of Sciences, and the Tianhe-1A platform at the National
Supercomputer Center in Tianjin for technical support and generous allocation
of CPU time.
\end{acknowledgments}


%

\setcounter{page}{1}
\setcounter{equation}{0}
\setcounter{figure}{0}
\renewcommand{\theequation}{S\arabic{equation}}
\renewcommand{\thefigure}{S\arabic{figure}}

\clearpage

\begin{widetext}
\section*{Supplemental Material}

\centerline{\bf Charge-Density-Wave Transitions of Dirac Fermions Coupled to Phonons}
\vskip3mm

\centerline{}

\vskip6mm

\subsection{SI.~Self-learning Monte Carlo}\label{sec:SLMC}

The self-learning Monte Carlo (SLMC) method was recently developed  to
propose efficient Monte Carlo updates in simulations of classical and
quantum many-body
systems~\cite{PhysRevB.95.041101,PhysRevB.95.241104,Xu2017SLMC,PhysRevB.98.041102,Nagaiself2017}. 
Its central idea is to make use of learning algorithms to construct
an approximate effective Hamiltonian that can be efficiently simulated
to guide the Monte Carlo simulation~\cite{PhysRevB.95.041101,PhysRevB.95.241104,Xu2017SLMC,LiHuang2017a,LiHuang2017b}.
SLMC has been shown to yield substantial improvements over traditional Monte
Carlo methods in terms of overcoming critical slowing-down and reducing
matrix operations in DQMC simulations. Exact simulations are ensured by evaluating the full fermion
determinant to decide about the acceptance of cumulative updates. However, 
this expensive operation can be done relatively infrequently, owing to the
accuracy of the learned effective action. For example, in 2D problems of
fermions coupled to critical bosonic fluctuations with itinerant quantum
critical points, linear lattice sizes $L$ up to 100 can be investigated at
high temperatures~\cite{Xu2017SLMC}. Low-temperature simulations of
antiferromagnetic Ising models coupled to fermions have been carried out
for triangular lattices with $L$ up to
48~\cite{PhysRevB.98.045116,2018arXiv180100127L} and square lattices with $L$
up to 60~\cite{2018arXiv180808878L}.

\begin{figure}[b]
\includegraphics[width=0.45\columnwidth]{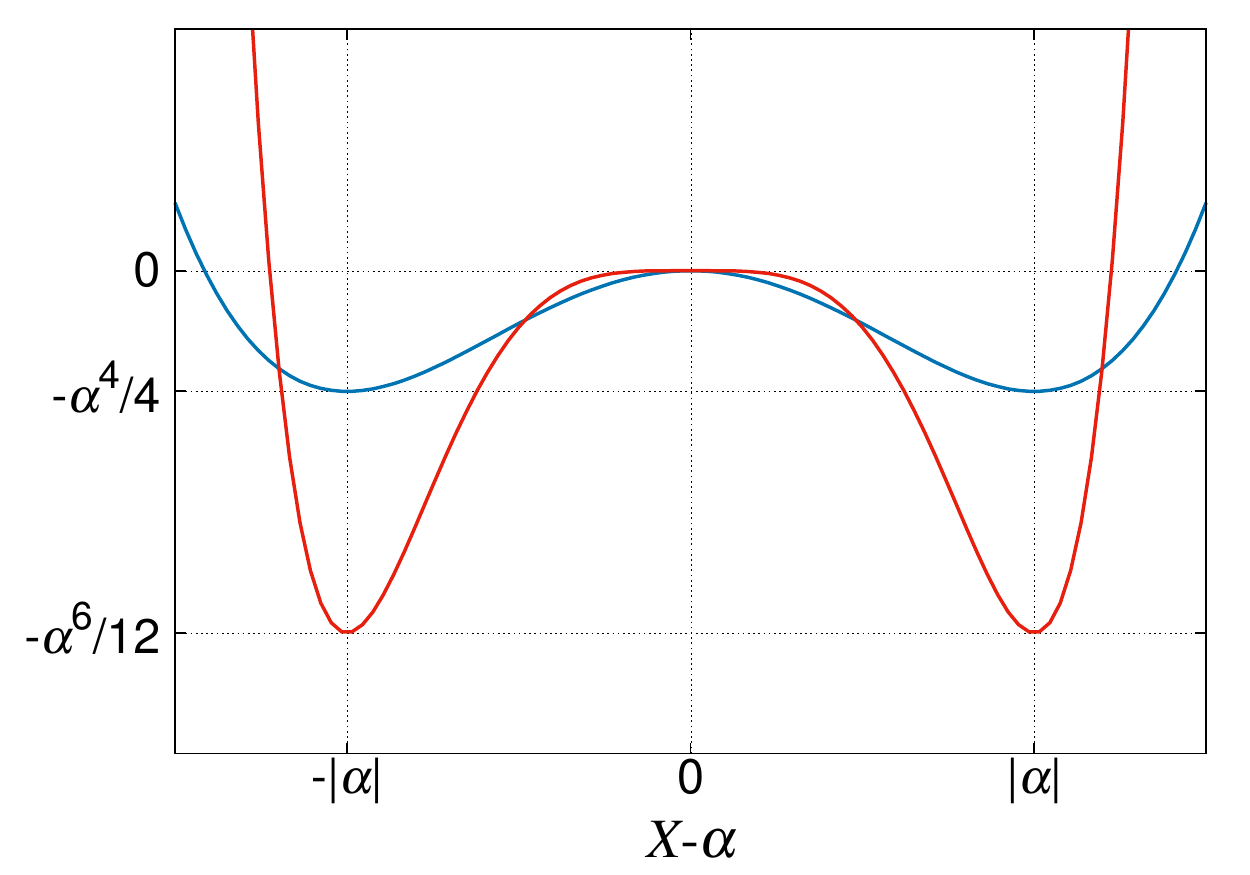}
\caption{\label{fig:S1}
The symmetric functions used to construct the phonon potential have
minima at $\pm |\alpha|$ with $\alpha=-{g}/{\Omega^2}$. The blue line
corresponds to $\frac{1}{4}(X-\alpha)^{4} - \frac {{\alpha}^2} {2} (X - \alpha )^2 $,
the red line to  $\frac{1}{6}(X-\alpha)^{6} - \frac {{\alpha}^2} {4} (X
- \alpha )^4 $. Figure reproduced from Ref.~\cite{PhysRevB.98.041102}.}
\end{figure}

SLMC has previously been successfully applied to the Holstein model on the square
lattice  \cite{PhysRevB.98.041102}. The approach used here is based on Ref.~\cite{PhysRevB.98.045116}. The first step
 is to obtain an effective model by self-learning on configurations generated with
DQMC. The model was chosen to be of the form (up to a constant)
\begin{align}
-\beta H^{\text{eff}} & = 
J_{k}\sum_{i\tau}(X_{i\tau+1}-X_{i\tau})^{2} 
\nonumber  
+ J_{p}\sum_{i\tau}\left(\frac{1}{4}(X_{i\tau}-\alpha)^{4} - \frac
{{\alpha}^2} {2} (X_{i\tau} - \alpha )^2 \right)    
\nonumber  
+ J_{p}'\sum_{i\tau}\left(\frac{1}{6}(X_{i\tau}-\alpha)^{6} - \frac
{{\alpha}^2} {4} (X_{i\tau} - \alpha )^4 \right)   
\nonumber  \\
 &+ J_{nn}\sum_{\langle ij \rangle
  \tau}(X_{i\tau}-\alpha)(X_{j\tau}-\alpha) 
+  J_{nn}'\sum_{i\langle \tau\tau' \rangle}(X_{i\tau}-\alpha
)(X_{i\tau'}-\alpha)\,.
\label{eq:effectiveHamHolsteinmain}
\end{align}
Here, the $J_k$ term comes from the phonon kinetic energy, the $J_p$ and
$J_p'$ terms are functions that produce the two global minima visible in
Fig.~\ref{fig:S1} (see below), while $J_{nn}$ and $J'_{nn}$ are nearest-neighbor
interactions in space and imaginary time, respectively. Importantly, the
effective model has a built-in global $Z_2$ symmetry related to invariance 
under a global mirror operation on $X$ with axis $\alpha$. The two potential
minima of the Holstein model are symmetric with respect to $X=- g/
{\Omega^2}\equiv \alpha$, see Fig.~\ref{fig:S1}. We find that for the phonon
fields in the Holstein model, two functions are sufficient to fit an
appropriate barrier width and height.

With the effective model in the form of Eq.~(\ref{eq:effectiveHamHolsteinmain}), the training procedure is
straight forward. Given a configuration $\mathcal{X}$ of the phonon fields and
a corresponding weight $\omega[\mathcal{X}]$, generated in DQMC, the learning
objective is 
\begin{equation}
  -\beta H^{\text{eff}} [\mathcal{X}] = \ln \left(  \omega[\mathcal{X}]
  \right)\,. 
  \label{eq:fitting}
\end{equation}
Combining Eqs.~(\ref{eq:effectiveHamHolsteinmain}) and~(\ref{eq:fitting}),
optimized values of $J_k$, $J_p$, $J_p'$, $J_{nn}$ and $J_{nn}'$ can be
readily obtained through multi-linear
regression~\cite{liu2016self,liu2016fermion,Xu2017SLMC} using the
configurations prepared with DQMC. For each temperature, we determined
$H^{\text{eff}}$ from $20,000$ configurations  for $L=6$, after which the
model was also used for larger $L$. The fitted parameters for the case
of $L=6$, $\beta=6$, and $\lambda=1/3$ are reported in Table~\ref{table1}.

\begin{table}[t]
    \caption{Fitted values of  $J_k$, $J_p$, $J_p'$, $J_{nn}$ and
      $J_{nn}'$  from a multi-linear regression~\cite{liu2016self,liu2016fermion,Xu2017SLMC} with $20,000$
      DQMC configurations.} 
    \centering 
    \begin{tabular}{c c c c c} 
    \hline\hline\\[-1ex] 
    $J_k$ & $J_p$ & $J_p'$ & $J_{nn}$ & $J_{nn}'$ \\ [0.5ex] 
    \hline\\[-1ex] 
     -5.00E1  &  1.61E-2  &  -6.07E-4  &  3.29E-2  &  8.88E-2  \\ [1ex] 
    \hline 
    \end{tabular}
    \label{table1} 
\end{table}
    
The effective model guides the Monte Carlo simulation of the original model
in terms of proposes updates of the phonon fields based on
Eq.~(\ref{eq:effectiveHamHolsteinmain}). This is the so-called cumulative
update in SLMC~\cite{liu2016fermion,Xu2017SLMC}. We then calculate the
acceptance ratio of the final phonon-field configuration via the expensive
fermion determinant only rarely.  

There are two advantages of SLMC over
DQMC. First, the effective model is purely bosonic and its local update is
$\mathcal{O}(1)$ since it bypasses the calculation of fermion determinants. Second,
since the effective model is bosonic, global updates such as cluster update
schemes~\cite{SwendsenWang1987,Wolff1989} or Hamiltonian Monte
Carlo~\cite{duane1987hybrid} are easy to implement.  This is crucial since
global updates in conventional DQMC actually worsen the scaling from
$O(N^3 L_\tau)$ to $O(N^4 L_\tau)$. Using the low-cost global updates in
combination with the SLMC approach allows for reliable, large-scale simulations.
In this work, we used the cumulative update (local update plus 10 block
updates) in the weak-coupling regime and the Wolff update in the strong
coupling regime.

\subsection{SII.~Analytic continuation of the single-particle Green function}

For the spectral functions shown in Fig.~2 of the main text, we measured the
single-particle Green function
\begin{equation}\label{eq:greenfunction}
  G^{A/B}_\sigma(\bk,\tau)= \las c^\dag_{\bk\sigma,A/B}(\tau)  c^\nag_{\bk\sigma,A/B}(0)\ras\,,
\end{equation}
averaged it over spin ($\sigma=\UP,\DO$) and sublattices ($A/B$), before using the ALF
implementation \cite{ALF17} of the so-called stochastic maximum entropy
method \cite{Beach04a} to extract the spectral function
\begin{align}\label{eqn:akw}
  A(\bk,\omega)&=-\pi^{-1}\mathrm{Im}\, G(\bk,\omega)\,.
\end{align}

\subsection{SIII.~Absence of superconductivity}

Figure~\ref{fig:S2} shows the correlation ratio for $s$-wave pairing,
%
  $R_\text{p} = 1 - {S_\text{p}(\boldsymbol{Q}+\delta\boldsymbol{Q})}/{S_\text{p}(\boldsymbol{Q})}$,
%
with
\begin{equation}
  S_\text{p} = \frac{1}{L^2}\sum_{ij} e^{-\rmi  \boldsymbol{q}\cdot(\boldsymbol{r}_i-\boldsymbol{r}_j)}\las
  \hat{\Delta}^\dag_{i,A} \hat{\Delta}^{}_{j,A} + \hat{\Delta}^\dag_{i,B} \hat{\Delta}^{}_{j,B}  \ras
\,,
\end{equation}
 $\boldsymbol{Q}=(0,0)$, and $\hat{\Delta}_{i,A/B}=c_{i\UP,A/B}c_{i\DO,A/B}$. Whereas the correlation
ratio for CDW order in Fig.~\ref{fig:S2}(a) shows a clear phase transition with
long-range order for $\lambda>\lambda_c\approx 0.2375$, $R_\text{p}$ in Fig.~\ref{fig:S2}(b)
remains small and decreases with increasing system size, suggesting the
absence of any significant pairing correlations for the parameters considered.

\begin{figure}
  \includegraphics[width=0.475\columnwidth]{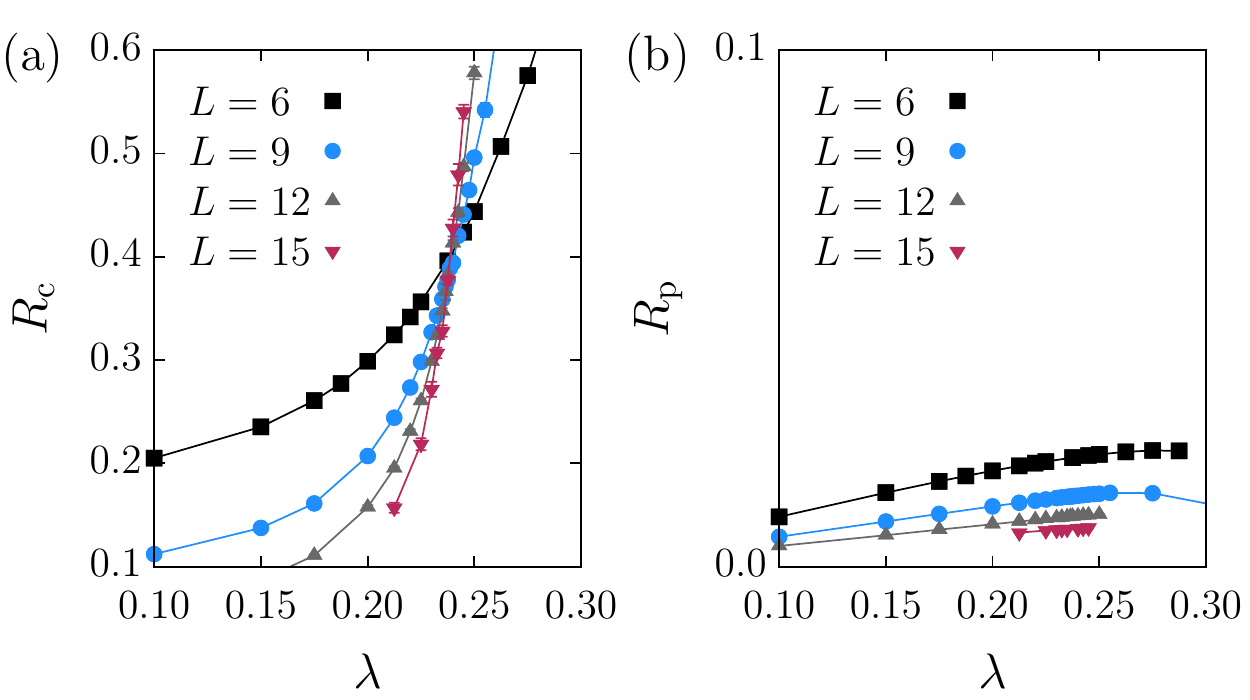}
  \caption{\label{fig:S2}
  Correlation ratios for (a) CDW order and (b) s-wave superconductivity from
  CT-INT simulations with $\beta t=L$.
  }
\end{figure}

\end{widetext}


\begin{thebibliography}{80}%
\makeatletter
\providecommand \@ifxundefined [1]{%
 \@ifx{#1\undefined}
}%
\providecommand \@ifnum [1]{%
 \ifnum #1\expandafter \@firstoftwo
 \else \expandafter \@secondoftwo
 \fi
}%
\providecommand \@ifx [1]{%
 \ifx #1\expandafter \@firstoftwo
 \else \expandafter \@secondoftwo
 \fi
}%
\providecommand \natexlab [1]{#1}%
\providecommand \enquote  [1]{``#1''}%
\providecommand \bibnamefont  [1]{#1}%
\providecommand \bibfnamefont [1]{#1}%
\providecommand \citenamefont [1]{#1}%
\providecommand \href@noop [0]{\@secondoftwo}%
\providecommand \href [0]{\begingroup \@sanitize@url \@href}%
\providecommand \@href[1]{\@@startlink{#1}\@@href}%
\providecommand \@@href[1]{\endgroup#1\@@endlink}%
\providecommand \@sanitize@url [0]{\catcode `\\12\catcode `\$12\catcode
  `\&12\catcode `\#12\catcode `\^12\catcode `\_12\catcode `\%12\relax}%
\providecommand \@@startlink[1]{}%
\providecommand \@@endlink[0]{}%
\providecommand \url  [0]{\begingroup\@sanitize@url \@url }%
\providecommand \@url [1]{\endgroup\@href {#1}{\urlprefix }}%
\providecommand \urlprefix  [0]{URL }%
\providecommand \Eprint [0]{\href }%
\providecommand \doibase [0]{http://dx.doi.org/}%
\providecommand \selectlanguage [0]{\@gobble}%
\providecommand \bibinfo  [0]{\@secondoftwo}%
\providecommand \bibfield  [0]{\@secondoftwo}%
\providecommand \translation [1]{[#1]}%
\providecommand \BibitemOpen [0]{}%
\providecommand \bibitemStop [0]{}%
\providecommand \bibitemNoStop [0]{.\EOS\space}%
\providecommand \EOS [0]{\spacefactor3000\relax}%
\providecommand \BibitemShut  [1]{\csname bibitem#1\endcsname}%
\let\auto@bib@innerbib\@empty
\bibitem [{\citenamefont {Neto}\ \emph {et~al.}(2009)\citenamefont {Neto},
  \citenamefont {Guinea}, \citenamefont {Peres}, \citenamefont {Novoselov},\
  and\ \citenamefont {Geim}}]{Neto_rev}%
  \BibitemOpen
  \bibfield  {author} {\bibinfo {author} {\bibfnamefont {A.~H.~C.}\
  \bibnamefont {Neto}}, \bibinfo {author} {\bibfnamefont {F.}~\bibnamefont
  {Guinea}}, \bibinfo {author} {\bibfnamefont {N.~M.~R.}\ \bibnamefont
  {Peres}}, \bibinfo {author} {\bibfnamefont {K.~S.}\ \bibnamefont
  {Novoselov}}, \ and\ \bibinfo {author} {\bibfnamefont {A.~K.}\ \bibnamefont
  {Geim}},\ }\href {\doibase 10.1103/RevModPhys.81.109} {\bibfield  {journal}
  {\bibinfo  {journal} {Rev. Mod. Phys.}\ }\textbf {\bibinfo {volume} {81}},\
  \bibinfo {eid} {109} (\bibinfo {year} {2009})}\BibitemShut {NoStop}%
\bibitem [{\citenamefont {Tang}\ \emph {et~al.}(2018)\citenamefont {Tang},
  \citenamefont {Leaw}, \citenamefont {Rodrigues}, \citenamefont {Herbut},
  \citenamefont {Sengupta}, \citenamefont {Assaad},\ and\ \citenamefont
  {Adam}}]{tang2018role}%
  \BibitemOpen
  \bibfield  {author} {\bibinfo {author} {\bibfnamefont {H.-K.}\ \bibnamefont
  {Tang}}, \bibinfo {author} {\bibfnamefont {J.}~\bibnamefont {Leaw}}, \bibinfo
  {author} {\bibfnamefont {J.}~\bibnamefont {Rodrigues}}, \bibinfo {author}
  {\bibfnamefont {I.}~\bibnamefont {Herbut}}, \bibinfo {author} {\bibfnamefont
  {P.}~\bibnamefont {Sengupta}}, \bibinfo {author} {\bibfnamefont
  {F.}~\bibnamefont {Assaad}}, \ and\ \bibinfo {author} {\bibfnamefont
  {S.}~\bibnamefont {Adam}},\ }\href@noop {} {\bibfield  {journal} {\bibinfo
  {journal} {Science}\ }\textbf {\bibinfo {volume} {361}},\ \bibinfo {pages}
  {570} (\bibinfo {year} {2018})}\BibitemShut {NoStop}%
\bibitem [{\citenamefont {Reis}\ \emph {et~al.}(2017)\citenamefont {Reis},
  \citenamefont {Li}, \citenamefont {Dudy}, \citenamefont {Bauernfeind},
  \citenamefont {Glass}, \citenamefont {Hanke}, \citenamefont {Thomale},
  \citenamefont {Sch{\"a}fer},\ and\ \citenamefont
  {Claessen}}]{reis2017bismuthene}%
  \BibitemOpen
  \bibfield  {author} {\bibinfo {author} {\bibfnamefont {F.}~\bibnamefont
  {Reis}}, \bibinfo {author} {\bibfnamefont {G.}~\bibnamefont {Li}}, \bibinfo
  {author} {\bibfnamefont {L.}~\bibnamefont {Dudy}}, \bibinfo {author}
  {\bibfnamefont {M.}~\bibnamefont {Bauernfeind}}, \bibinfo {author}
  {\bibfnamefont {S.}~\bibnamefont {Glass}}, \bibinfo {author} {\bibfnamefont
  {W.}~\bibnamefont {Hanke}}, \bibinfo {author} {\bibfnamefont
  {R.}~\bibnamefont {Thomale}}, \bibinfo {author} {\bibfnamefont
  {J.}~\bibnamefont {Sch{\"a}fer}}, \ and\ \bibinfo {author} {\bibfnamefont
  {R.}~\bibnamefont {Claessen}},\ }\href@noop {} {\bibfield  {journal}
  {\bibinfo  {journal} {Science}\ }\textbf {\bibinfo {volume} {357}},\ \bibinfo
  {pages} {287} (\bibinfo {year} {2017})}\BibitemShut {NoStop}%
\bibitem [{\citenamefont {Cao}\ \emph {et~al.}(2018)\citenamefont {Cao},
  \citenamefont {Fatemi}, \citenamefont {Fang}, \citenamefont {Watanabe},
  \citenamefont {Taniguchi}, \citenamefont {Kaxiras},\ and\ \citenamefont
  {Jarillo-Herrero}}]{cao2018unconventional}%
  \BibitemOpen
  \bibfield  {author} {\bibinfo {author} {\bibfnamefont {Y.}~\bibnamefont
  {Cao}}, \bibinfo {author} {\bibfnamefont {V.}~\bibnamefont {Fatemi}},
  \bibinfo {author} {\bibfnamefont {S.}~\bibnamefont {Fang}}, \bibinfo {author}
  {\bibfnamefont {K.}~\bibnamefont {Watanabe}}, \bibinfo {author}
  {\bibfnamefont {T.}~\bibnamefont {Taniguchi}}, \bibinfo {author}
  {\bibfnamefont {E.}~\bibnamefont {Kaxiras}}, \ and\ \bibinfo {author}
  {\bibfnamefont {P.}~\bibnamefont {Jarillo-Herrero}},\ }\href@noop {}
  {\bibfield  {journal} {\bibinfo  {journal} {Nature}\ }\textbf {\bibinfo
  {volume} {556}},\ \bibinfo {pages} {43} (\bibinfo {year} {2018})}\BibitemShut
  {NoStop}%
\bibitem [{\citenamefont {Manzeli}\ \emph {et~al.}(2017)\citenamefont
  {Manzeli}, \citenamefont {Ovchinnikov}, \citenamefont {Pasquier},
  \citenamefont {Yazyev},\ and\ \citenamefont {Kis}}]{manzeli20172d}%
  \BibitemOpen
  \bibfield  {author} {\bibinfo {author} {\bibfnamefont {S.}~\bibnamefont
  {Manzeli}}, \bibinfo {author} {\bibfnamefont {D.}~\bibnamefont
  {Ovchinnikov}}, \bibinfo {author} {\bibfnamefont {D.}~\bibnamefont
  {Pasquier}}, \bibinfo {author} {\bibfnamefont {O.~V.}\ \bibnamefont
  {Yazyev}}, \ and\ \bibinfo {author} {\bibfnamefont {A.}~\bibnamefont {Kis}},\
  }\href@noop {} {\bibfield  {journal} {\bibinfo  {journal} {Nat. Rev.
  Materials}\ }\textbf {\bibinfo {volume} {2}},\ \bibinfo {pages} {17033}
  (\bibinfo {year} {2017})}\BibitemShut {NoStop}%
\bibitem [{\citenamefont {Semenoff}(1984)}]{PhysRevLett.53.2449}%
  \BibitemOpen
  \bibfield  {author} {\bibinfo {author} {\bibfnamefont {G.~W.}\ \bibnamefont
  {Semenoff}},\ }\href {\doibase 10.1103/PhysRevLett.53.2449} {\bibfield
  {journal} {\bibinfo  {journal} {Phys. Rev. Lett.}\ }\textbf {\bibinfo
  {volume} {53}},\ \bibinfo {pages} {2449} (\bibinfo {year}
  {1984})}\BibitemShut {NoStop}%
\bibitem [{\citenamefont {Haldane}(1988)}]{Haldane98}%
  \BibitemOpen
  \bibfield  {author} {\bibinfo {author} {\bibfnamefont {F.~D.~M.}\
  \bibnamefont {Haldane}},\ }\href {\doibase 10.1103/PhysRevLett.61.2015}
  {\bibfield  {journal} {\bibinfo  {journal} {Phys. Rev. Lett.}\ }\textbf
  {\bibinfo {volume} {61}},\ \bibinfo {pages} {2015} (\bibinfo {year}
  {1988})}\BibitemShut {NoStop}%
\bibitem [{\citenamefont {Herbut}(2006)}]{PhysRevLett.97.146401}%
  \BibitemOpen
  \bibfield  {author} {\bibinfo {author} {\bibfnamefont {I.~F.}\ \bibnamefont
  {Herbut}},\ }\href {\doibase 10.1103/PhysRevLett.97.146401} {\bibfield
  {journal} {\bibinfo  {journal} {Phys. Rev. Lett.}\ }\textbf {\bibinfo
  {volume} {97}},\ \bibinfo {pages} {146401} (\bibinfo {year}
  {2006})}\BibitemShut {NoStop}%
\bibitem [{\citenamefont {Herbut}\ \emph
  {et~al.}(2009{\natexlab{a}})\citenamefont {Herbut}, \citenamefont
  {Juri\ifmmode \check{c}\else \v{c}\fi{}i\ifmmode~\acute{c}\else \'{c}\fi{}},\
  and\ \citenamefont {Vafek}}]{Herbut09a}%
  \BibitemOpen
  \bibfield  {author} {\bibinfo {author} {\bibfnamefont {I.~F.}\ \bibnamefont
  {Herbut}}, \bibinfo {author} {\bibfnamefont {V.}~\bibnamefont {Juri\ifmmode
  \check{c}\else \v{c}\fi{}i\ifmmode~\acute{c}\else \'{c}\fi{}}}, \ and\
  \bibinfo {author} {\bibfnamefont {O.}~\bibnamefont {Vafek}},\ }\href
  {\doibase 10.1103/PhysRevB.80.075432} {\bibfield  {journal} {\bibinfo
  {journal} {Phys. Rev. B}\ }\textbf {\bibinfo {volume} {80}},\ \bibinfo
  {pages} {075432} (\bibinfo {year} {2009}{\natexlab{a}})}\BibitemShut
  {NoStop}%
\bibitem [{\citenamefont {Scherer}\ and\ \citenamefont
  {Herbut}(2016)}]{scherer2016gauge}%
  \BibitemOpen
  \bibfield  {author} {\bibinfo {author} {\bibfnamefont {M.~M.}\ \bibnamefont
  {Scherer}}\ and\ \bibinfo {author} {\bibfnamefont {I.~F.}\ \bibnamefont
  {Herbut}},\ }\href {\doibase 10.1103/PhysRevB.94.205136} {\bibfield
  {journal} {\bibinfo  {journal} {Phys. Rev. B}\ }\textbf {\bibinfo {volume}
  {94}},\ \bibinfo {pages} {205136} (\bibinfo {year} {2016})}\BibitemShut
  {NoStop}%
\bibitem [{\citenamefont {Zhou}\ \emph {et~al.}(2016)\citenamefont {Zhou},
  \citenamefont {Wang}, \citenamefont {Meng}, \citenamefont {Wang},\ and\
  \citenamefont {Wu}}]{zhou2016mott}%
  \BibitemOpen
  \bibfield  {author} {\bibinfo {author} {\bibfnamefont {Z.}~\bibnamefont
  {Zhou}}, \bibinfo {author} {\bibfnamefont {D.}~\bibnamefont {Wang}}, \bibinfo
  {author} {\bibfnamefont {Z.~Y.}\ \bibnamefont {Meng}}, \bibinfo {author}
  {\bibfnamefont {Y.}~\bibnamefont {Wang}}, \ and\ \bibinfo {author}
  {\bibfnamefont {C.}~\bibnamefont {Wu}},\ }\href {\doibase
  10.1103/PhysRevB.93.245157} {\bibfield  {journal} {\bibinfo  {journal} {Phys.
  Rev. B}\ }\textbf {\bibinfo {volume} {93}},\ \bibinfo {pages} {245157}
  (\bibinfo {year} {2016})}\BibitemShut {NoStop}%
\bibitem [{\citenamefont {Li}\ \emph {et~al.}(2017)\citenamefont {Li},
  \citenamefont {Jiang}, \citenamefont {Jian},\ and\ \citenamefont
  {Yao}}]{li2017fermion}%
  \BibitemOpen
  \bibfield  {author} {\bibinfo {author} {\bibfnamefont {Z.-X.}\ \bibnamefont
  {Li}}, \bibinfo {author} {\bibfnamefont {Y.-F.}\ \bibnamefont {Jiang}},
  \bibinfo {author} {\bibfnamefont {S.-K.}\ \bibnamefont {Jian}}, \ and\
  \bibinfo {author} {\bibfnamefont {H.}~\bibnamefont {Yao}},\ }\href@noop {}
  {\bibfield  {journal} {\bibinfo  {journal} {Nature Comm.}\ }\textbf {\bibinfo
  {volume} {8}},\ \bibinfo {pages} {314} (\bibinfo {year} {2017})}\BibitemShut
  {NoStop}%
\bibitem [{\citenamefont {Classen}\ \emph {et~al.}(2017)\citenamefont
  {Classen}, \citenamefont {Herbut},\ and\ \citenamefont
  {Scherer}}]{classen2017fluctuation}%
  \BibitemOpen
  \bibfield  {author} {\bibinfo {author} {\bibfnamefont {L.}~\bibnamefont
  {Classen}}, \bibinfo {author} {\bibfnamefont {I.~F.}\ \bibnamefont {Herbut}},
  \ and\ \bibinfo {author} {\bibfnamefont {M.~M.}\ \bibnamefont {Scherer}},\
  }\href {\doibase 10.1103/PhysRevB.96.115132} {\bibfield  {journal} {\bibinfo
  {journal} {Phys. Rev. B}\ }\textbf {\bibinfo {volume} {96}},\ \bibinfo
  {pages} {115132} (\bibinfo {year} {2017})}\BibitemShut {NoStop}%
\bibitem [{\citenamefont {He}\ \emph {et~al.}(2018)\citenamefont {He},
  \citenamefont {Xu}, \citenamefont {Sun}, \citenamefont {Assaad},
  \citenamefont {Meng},\ and\ \citenamefont {Lu}}]{he2018dyn}%
  \BibitemOpen
  \bibfield  {author} {\bibinfo {author} {\bibfnamefont {Y.-Y.}\ \bibnamefont
  {He}}, \bibinfo {author} {\bibfnamefont {X.~Y.}\ \bibnamefont {Xu}}, \bibinfo
  {author} {\bibfnamefont {K.}~\bibnamefont {Sun}}, \bibinfo {author}
  {\bibfnamefont {F.~F.}\ \bibnamefont {Assaad}}, \bibinfo {author}
  {\bibfnamefont {Z.~Y.}\ \bibnamefont {Meng}}, \ and\ \bibinfo {author}
  {\bibfnamefont {Z.-Y.}\ \bibnamefont {Lu}},\ }\href {\doibase
  10.1103/PhysRevB.97.081110} {\bibfield  {journal} {\bibinfo  {journal} {Phys.
  Rev. B}\ }\textbf {\bibinfo {volume} {97}},\ \bibinfo {pages} {081110}
  (\bibinfo {year} {2018})}\BibitemShut {NoStop}%
\bibitem [{\citenamefont {Torres}\ \emph {et~al.}(2018)\citenamefont {Torres},
  \citenamefont {Classen}, \citenamefont {Herbut},\ and\ \citenamefont
  {Scherer}}]{torres2018fermion}%
  \BibitemOpen
  \bibfield  {author} {\bibinfo {author} {\bibfnamefont {E.}~\bibnamefont
  {Torres}}, \bibinfo {author} {\bibfnamefont {L.}~\bibnamefont {Classen}},
  \bibinfo {author} {\bibfnamefont {I.~F.}\ \bibnamefont {Herbut}}, \ and\
  \bibinfo {author} {\bibfnamefont {M.~M.}\ \bibnamefont {Scherer}},\ }\href
  {\doibase 10.1103/PhysRevB.97.125137} {\bibfield  {journal} {\bibinfo
  {journal} {Phys. Rev. B}\ }\textbf {\bibinfo {volume} {97}},\ \bibinfo
  {pages} {125137} (\bibinfo {year} {2018})}\BibitemShut {NoStop}%
\bibitem [{\citenamefont {Xu}\ \emph {et~al.}(2018)\citenamefont {Xu},
  \citenamefont {Law},\ and\ \citenamefont {Lee}}]{xu2018kekule}%
  \BibitemOpen
  \bibfield  {author} {\bibinfo {author} {\bibfnamefont {X.~Y.}\ \bibnamefont
  {Xu}}, \bibinfo {author} {\bibfnamefont {K.~T.}\ \bibnamefont {Law}}, \ and\
  \bibinfo {author} {\bibfnamefont {P.~A.}\ \bibnamefont {Lee}},\ }\href
  {\doibase 10.1103/PhysRevB.98.121406} {\bibfield  {journal} {\bibinfo
  {journal} {Phys. Rev. B}\ }\textbf {\bibinfo {volume} {98}},\ \bibinfo
  {pages} {121406} (\bibinfo {year} {2018})}\BibitemShut {NoStop}%
\bibitem [{\citenamefont {Lang}\ and\ \citenamefont
  {L{\"a}uchli}(2018)}]{lang2018quantum}%
  \BibitemOpen
  \bibfield  {author} {\bibinfo {author} {\bibfnamefont {T.~C.}\ \bibnamefont
  {Lang}}\ and\ \bibinfo {author} {\bibfnamefont {A.~M.}\ \bibnamefont
  {L{\"a}uchli}},\ }\href@noop {} {\bibfield  {journal} {\bibinfo  {journal}
  {arXiv preprint arXiv:1808.01230}\ } (\bibinfo {year} {2018})}\BibitemShut
  {NoStop}%
\bibitem [{\citenamefont {Senthil}\ \emph {et~al.}(2004)\citenamefont
  {Senthil}, \citenamefont {Vishwanath}, \citenamefont {Balents}, \citenamefont
  {Sachdev},\ and\ \citenamefont {Fisher}}]{senthil2004decon}%
  \BibitemOpen
  \bibfield  {author} {\bibinfo {author} {\bibfnamefont {T.}~\bibnamefont
  {Senthil}}, \bibinfo {author} {\bibfnamefont {A.}~\bibnamefont {Vishwanath}},
  \bibinfo {author} {\bibfnamefont {L.}~\bibnamefont {Balents}}, \bibinfo
  {author} {\bibfnamefont {S.}~\bibnamefont {Sachdev}}, \ and\ \bibinfo
  {author} {\bibfnamefont {M.~P.~A.}\ \bibnamefont {Fisher}},\ }\href {\doibase
  10.1126/science.1091806} {\bibfield  {journal} {\bibinfo  {journal}
  {Science}\ }\textbf {\bibinfo {volume} {303}},\ \bibinfo {pages} {1490}
  (\bibinfo {year} {2004})}\BibitemShut {NoStop}%
\bibitem [{\citenamefont {Sato}\ \emph {et~al.}(2017)\citenamefont {Sato},
  \citenamefont {Hohenadler},\ and\ \citenamefont {Assaad}}]{sato2017dirac}%
  \BibitemOpen
  \bibfield  {author} {\bibinfo {author} {\bibfnamefont {T.}~\bibnamefont
  {Sato}}, \bibinfo {author} {\bibfnamefont {M.}~\bibnamefont {Hohenadler}}, \
  and\ \bibinfo {author} {\bibfnamefont {F.~F.}\ \bibnamefont {Assaad}},\
  }\href {\doibase 10.1103/PhysRevLett.119.197203} {\bibfield  {journal}
  {\bibinfo  {journal} {Phys. Rev. Lett.}\ }\textbf {\bibinfo {volume} {119}},\
  \bibinfo {pages} {197203} (\bibinfo {year} {2017})}\BibitemShut {NoStop}%
\bibitem [{\citenamefont {Qin}\ \emph {et~al.}(2017)\citenamefont {Qin},
  \citenamefont {He}, \citenamefont {You}, \citenamefont {Lu}, \citenamefont
  {Sen}, \citenamefont {Sandvik}, \citenamefont {Xu},\ and\ \citenamefont
  {Meng}}]{qin2017dual}%
  \BibitemOpen
  \bibfield  {author} {\bibinfo {author} {\bibfnamefont {Y.~Q.}\ \bibnamefont
  {Qin}}, \bibinfo {author} {\bibfnamefont {Y.-Y.}\ \bibnamefont {He}},
  \bibinfo {author} {\bibfnamefont {Y.-Z.}\ \bibnamefont {You}}, \bibinfo
  {author} {\bibfnamefont {Z.-Y.}\ \bibnamefont {Lu}}, \bibinfo {author}
  {\bibfnamefont {A.}~\bibnamefont {Sen}}, \bibinfo {author} {\bibfnamefont
  {A.~W.}\ \bibnamefont {Sandvik}}, \bibinfo {author} {\bibfnamefont
  {C.}~\bibnamefont {Xu}}, \ and\ \bibinfo {author} {\bibfnamefont {Z.~Y.}\
  \bibnamefont {Meng}},\ }\href {\doibase 10.1103/PhysRevX.7.031052} {\bibfield
   {journal} {\bibinfo  {journal} {Phys. Rev. X}\ }\textbf {\bibinfo {volume}
  {7}},\ \bibinfo {pages} {031052} (\bibinfo {year} {2017})}\BibitemShut
  {NoStop}%
\bibitem [{\citenamefont {Wang}\ \emph {et~al.}(2017)\citenamefont {Wang},
  \citenamefont {Nahum}, \citenamefont {Metlitski}, \citenamefont {Xu},\ and\
  \citenamefont {Senthil}}]{wang2017decon}%
  \BibitemOpen
  \bibfield  {author} {\bibinfo {author} {\bibfnamefont {C.}~\bibnamefont
  {Wang}}, \bibinfo {author} {\bibfnamefont {A.}~\bibnamefont {Nahum}},
  \bibinfo {author} {\bibfnamefont {M.~A.}\ \bibnamefont {Metlitski}}, \bibinfo
  {author} {\bibfnamefont {C.}~\bibnamefont {Xu}}, \ and\ \bibinfo {author}
  {\bibfnamefont {T.}~\bibnamefont {Senthil}},\ }\href {\doibase
  10.1103/PhysRevX.7.031051} {\bibfield  {journal} {\bibinfo  {journal} {Phys.
  Rev. X}\ }\textbf {\bibinfo {volume} {7}},\ \bibinfo {pages} {031051}
  (\bibinfo {year} {2017})}\BibitemShut {NoStop}%
\bibitem [{\citenamefont {Sorella}\ and\ \citenamefont
  {Tosatti}(1992)}]{Sorella92}%
  \BibitemOpen
  \bibfield  {author} {\bibinfo {author} {\bibfnamefont {S.}~\bibnamefont
  {Sorella}}\ and\ \bibinfo {author} {\bibfnamefont {E.}~\bibnamefont
  {Tosatti}},\ }\href@noop {} {\bibfield  {journal} {\bibinfo  {journal}
  {Europhys. Lett.}\ }\textbf {\bibinfo {volume} {19}},\ \bibinfo {pages} {699}
  (\bibinfo {year} {1992})}\BibitemShut {NoStop}%
\bibitem [{\citenamefont {Sorella}\ \emph {et~al.}(2012)\citenamefont
  {Sorella}, \citenamefont {Otsuka},\ and\ \citenamefont {Yunoki}}]{Sorella12}%
  \BibitemOpen
  \bibfield  {author} {\bibinfo {author} {\bibfnamefont {S.}~\bibnamefont
  {Sorella}}, \bibinfo {author} {\bibfnamefont {Y.}~\bibnamefont {Otsuka}}, \
  and\ \bibinfo {author} {\bibfnamefont {S.}~\bibnamefont {Yunoki}},\ }\href
  {\doibase http://dx.doi.org/10.1038/srep00992} {\bibfield  {journal}
  {\bibinfo  {journal} {Sci. Rep.}\ }\textbf {\bibinfo {volume} {2}},\ \bibinfo
  {pages} {992} (\bibinfo {year} {2012})}\BibitemShut {NoStop}%
\bibitem [{\citenamefont {Assaad}\ and\ \citenamefont
  {Herbut}(2013)}]{Assaad13}%
  \BibitemOpen
  \bibfield  {author} {\bibinfo {author} {\bibfnamefont {F.~F.}\ \bibnamefont
  {Assaad}}\ and\ \bibinfo {author} {\bibfnamefont {I.~F.}\ \bibnamefont
  {Herbut}},\ }\href {\doibase 10.1103/PhysRevX.3.031010} {\bibfield  {journal}
  {\bibinfo  {journal} {Phys. Rev. X}\ }\textbf {\bibinfo {volume} {3}},\
  \bibinfo {pages} {031010} (\bibinfo {year} {2013})}\BibitemShut {NoStop}%
\bibitem [{\citenamefont {Hohenadler}\ \emph {et~al.}(2014)\citenamefont
  {Hohenadler}, \citenamefont {Parisen~Toldin}, \citenamefont {Herbut},\ and\
  \citenamefont {Assaad}}]{PhysRevB.90.085146}%
  \BibitemOpen
  \bibfield  {author} {\bibinfo {author} {\bibfnamefont {M.}~\bibnamefont
  {Hohenadler}}, \bibinfo {author} {\bibfnamefont {F.}~\bibnamefont
  {Parisen~Toldin}}, \bibinfo {author} {\bibfnamefont {I.~F.}\ \bibnamefont
  {Herbut}}, \ and\ \bibinfo {author} {\bibfnamefont {F.~F.}\ \bibnamefont
  {Assaad}},\ }\href {\doibase 10.1103/PhysRevB.90.085146} {\bibfield
  {journal} {\bibinfo  {journal} {Phys. Rev. B}\ }\textbf {\bibinfo {volume}
  {90}},\ \bibinfo {pages} {085146} (\bibinfo {year} {2014})}\BibitemShut
  {NoStop}%
\bibitem [{\citenamefont {Raghu}\ \emph {et~al.}(2008)\citenamefont {Raghu},
  \citenamefont {Qi}, \citenamefont {Honerkamp},\ and\ \citenamefont
  {Zhang}}]{Raghu08}%
  \BibitemOpen
  \bibfield  {author} {\bibinfo {author} {\bibfnamefont {S.}~\bibnamefont
  {Raghu}}, \bibinfo {author} {\bibfnamefont {X.-L.}\ \bibnamefont {Qi}},
  \bibinfo {author} {\bibfnamefont {C.}~\bibnamefont {Honerkamp}}, \ and\
  \bibinfo {author} {\bibfnamefont {S.-C.}\ \bibnamefont {Zhang}},\ }\href
  {\doibase 10.1103/PhysRevLett.100.156401} {\bibfield  {journal} {\bibinfo
  {journal} {Phys. Rev. Lett.}\ }\textbf {\bibinfo {volume} {100}},\ \bibinfo
  {eid} {156401} (\bibinfo {year} {2008})}\BibitemShut {NoStop}%
\bibitem [{\citenamefont {Huffman}\ and\ \citenamefont
  {Chandrasekharan}(2014)}]{PhysRevB.89.111101}%
  \BibitemOpen
  \bibfield  {author} {\bibinfo {author} {\bibfnamefont {E.~F.}\ \bibnamefont
  {Huffman}}\ and\ \bibinfo {author} {\bibfnamefont {S.}~\bibnamefont
  {Chandrasekharan}},\ }\href {\doibase 10.1103/PhysRevB.89.111101} {\bibfield
  {journal} {\bibinfo  {journal} {Phys. Rev. B}\ }\textbf {\bibinfo {volume}
  {89}},\ \bibinfo {pages} {111101} (\bibinfo {year} {2014})}\BibitemShut
  {NoStop}%
\bibitem [{\citenamefont {Wang}\ \emph {et~al.}(2014)\citenamefont {Wang},
  \citenamefont {Corboz},\ and\ \citenamefont
  {Troyer}}]{1367-2630-16-10-103008}%
  \BibitemOpen
  \bibfield  {author} {\bibinfo {author} {\bibfnamefont {L.}~\bibnamefont
  {Wang}}, \bibinfo {author} {\bibfnamefont {P.}~\bibnamefont {Corboz}}, \ and\
  \bibinfo {author} {\bibfnamefont {M.}~\bibnamefont {Troyer}},\ }\href
  {http://stacks.iop.org/1367-2630/16/i=10/a=103008} {\bibfield  {journal}
  {\bibinfo  {journal} {New J. Phys.}\ }\textbf {\bibinfo {volume} {16}},\
  \bibinfo {pages} {103008} (\bibinfo {year} {2014})}\BibitemShut {NoStop}%
\bibitem [{\citenamefont {Li}\ \emph {et~al.}(2015)\citenamefont {Li},
  \citenamefont {Jiang},\ and\ \citenamefont {Yao}}]{li2015fermion}%
  \BibitemOpen
  \bibfield  {author} {\bibinfo {author} {\bibfnamefont {Z.-X.}\ \bibnamefont
  {Li}}, \bibinfo {author} {\bibfnamefont {Y.-F.}\ \bibnamefont {Jiang}}, \
  and\ \bibinfo {author} {\bibfnamefont {H.}~\bibnamefont {Yao}},\ }\href
  {http://stacks.iop.org/1367-2630/17/i=8/a=085003} {\bibfield  {journal}
  {\bibinfo  {journal} {New J. Phys.}\ }\textbf {\bibinfo {volume} {17}},\
  \bibinfo {pages} {085003} (\bibinfo {year} {2015})}\BibitemShut {NoStop}%
\bibitem [{\citenamefont {Hesselmann}\ and\ \citenamefont
  {Wessel}(2016)}]{Wessel2016}%
  \BibitemOpen
  \bibfield  {author} {\bibinfo {author} {\bibfnamefont {S.}~\bibnamefont
  {Hesselmann}}\ and\ \bibinfo {author} {\bibfnamefont {S.}~\bibnamefont
  {Wessel}},\ }\href {\doibase 10.1103/PhysRevB.93.155157} {\bibfield
  {journal} {\bibinfo  {journal} {Phys. Rev. B}\ }\textbf {\bibinfo {volume}
  {93}},\ \bibinfo {pages} {155157} (\bibinfo {year} {2016})}\BibitemShut
  {NoStop}%
\bibitem [{\citenamefont {Capponi}(2016)}]{capponi2016phase}%
  \BibitemOpen
  \bibfield  {author} {\bibinfo {author} {\bibfnamefont {S.}~\bibnamefont
  {Capponi}},\ }\href@noop {} {\bibfield  {journal} {\bibinfo  {journal} {J.
  Phys.: Condens. Matter}\ }\textbf {\bibinfo {volume} {29}},\ \bibinfo {pages}
  {043002} (\bibinfo {year} {2017})}\BibitemShut {NoStop}%
\bibitem [{\citenamefont {de~la Pe\~na}\ \emph {et~al.}(2017)\citenamefont
  {de~la Pe\~na}, \citenamefont {Lichtenstein},\ and\ \citenamefont
  {Honerkamp}}]{PhysRevB.95.085143}%
  \BibitemOpen
  \bibfield  {author} {\bibinfo {author} {\bibfnamefont {D.~S.}\ \bibnamefont
  {de~la Pe\~na}}, \bibinfo {author} {\bibfnamefont {J.}~\bibnamefont
  {Lichtenstein}}, \ and\ \bibinfo {author} {\bibfnamefont {C.}~\bibnamefont
  {Honerkamp}},\ }\href {\doibase 10.1103/PhysRevB.95.085143} {\bibfield
  {journal} {\bibinfo  {journal} {Phys. Rev. B}\ }\textbf {\bibinfo {volume}
  {95}},\ \bibinfo {pages} {085143} (\bibinfo {year} {2017})}\BibitemShut
  {NoStop}%
\bibitem [{\citenamefont {Bijelic}\ \emph {et~al.}(2018)\citenamefont
  {Bijelic}, \citenamefont {Kaneko}, \citenamefont {Gros},\ and\ \citenamefont
  {Valent\'{\i}}}]{PhysRevB.97.125142}%
  \BibitemOpen
  \bibfield  {author} {\bibinfo {author} {\bibfnamefont {M.}~\bibnamefont
  {Bijelic}}, \bibinfo {author} {\bibfnamefont {R.}~\bibnamefont {Kaneko}},
  \bibinfo {author} {\bibfnamefont {C.}~\bibnamefont {Gros}}, \ and\ \bibinfo
  {author} {\bibfnamefont {R.}~\bibnamefont {Valent\'{\i}}},\ }\href {\doibase
  10.1103/PhysRevB.97.125142} {\bibfield  {journal} {\bibinfo  {journal} {Phys.
  Rev. B}\ }\textbf {\bibinfo {volume} {97}},\ \bibinfo {pages} {125142}
  (\bibinfo {year} {2018})}\BibitemShut {NoStop}%
\bibitem [{\citenamefont {Chandrasekharan}\ and\ \citenamefont
  {Li}(2013)}]{chandrasekharan2013}%
  \BibitemOpen
  \bibfield  {author} {\bibinfo {author} {\bibfnamefont {S.}~\bibnamefont
  {Chandrasekharan}}\ and\ \bibinfo {author} {\bibfnamefont {A.}~\bibnamefont
  {Li}},\ }\href {\doibase 10.1103/PhysRevD.88.021701} {\bibfield  {journal}
  {\bibinfo  {journal} {Phys. Rev. D}\ }\textbf {\bibinfo {volume} {88}},\
  \bibinfo {pages} {021701} (\bibinfo {year} {2013})}\BibitemShut {NoStop}%
\bibitem [{\citenamefont {Weber}\ and\ \citenamefont
  {Hohenadler}(2018)}]{PhysRevB.98.085405}%
  \BibitemOpen
  \bibfield  {author} {\bibinfo {author} {\bibfnamefont {M.}~\bibnamefont
  {Weber}}\ and\ \bibinfo {author} {\bibfnamefont {M.}~\bibnamefont
  {Hohenadler}},\ }\href {\doibase 10.1103/PhysRevB.98.085405} {\bibfield
  {journal} {\bibinfo  {journal} {Phys. Rev. B}\ }\textbf {\bibinfo {volume}
  {98}},\ \bibinfo {pages} {085405} (\bibinfo {year} {2018})}\BibitemShut
  {NoStop}%
\bibitem [{\citenamefont {Chen}\ \emph {et~al.}(2018)\citenamefont {Chen},
  \citenamefont {Xu}, \citenamefont {Liu}, \citenamefont {Batrouni},
  \citenamefont {Scalettar},\ and\ \citenamefont {Meng}}]{PhysRevB.98.041102}%
  \BibitemOpen
  \bibfield  {author} {\bibinfo {author} {\bibfnamefont {C.}~\bibnamefont
  {Chen}}, \bibinfo {author} {\bibfnamefont {X.~Y.}\ \bibnamefont {Xu}},
  \bibinfo {author} {\bibfnamefont {J.}~\bibnamefont {Liu}}, \bibinfo {author}
  {\bibfnamefont {G.}~\bibnamefont {Batrouni}}, \bibinfo {author}
  {\bibfnamefont {R.}~\bibnamefont {Scalettar}}, \ and\ \bibinfo {author}
  {\bibfnamefont {Z.~Y.}\ \bibnamefont {Meng}},\ }\href {\doibase
  10.1103/PhysRevB.98.041102} {\bibfield  {journal} {\bibinfo  {journal} {Phys.
  Rev. B}\ }\textbf {\bibinfo {volume} {98}},\ \bibinfo {pages} {041102}
  (\bibinfo {year} {2018})}\BibitemShut {NoStop}%
\bibitem [{\citenamefont {Batrouni}\ and\ \citenamefont
  {Scalettar}(2019)}]{BaSc2018}%
  \BibitemOpen
  \bibfield  {author} {\bibinfo {author} {\bibfnamefont {G.~G.}\ \bibnamefont
  {Batrouni}}\ and\ \bibinfo {author} {\bibfnamefont {R.~T.}\ \bibnamefont
  {Scalettar}},\ }\href {\doibase 10.1103/PhysRevB.99.035114} {\bibfield
  {journal} {\bibinfo  {journal} {Phys. Rev. B}\ }\textbf {\bibinfo {volume}
  {99}},\ \bibinfo {pages} {035114} (\bibinfo {year} {2019})}\BibitemShut
  {NoStop}%
\bibitem [{\citenamefont {Karakuzu}\ \emph {et~al.}(2018)\citenamefont
  {Karakuzu}, \citenamefont {Seki},\ and\ \citenamefont
  {Sorella}}]{Ka.Se.So.18}%
  \BibitemOpen
  \bibfield  {author} {\bibinfo {author} {\bibfnamefont {S.}~\bibnamefont
  {Karakuzu}}, \bibinfo {author} {\bibfnamefont {K.}~\bibnamefont {Seki}}, \
  and\ \bibinfo {author} {\bibfnamefont {S.}~\bibnamefont {Sorella}},\ }\href
  {\doibase 10.1103/PhysRevB.98.201108} {\bibfield  {journal} {\bibinfo
  {journal} {Phys. Rev. B}\ }\textbf {\bibinfo {volume} {98}},\ \bibinfo
  {pages} {201108} (\bibinfo {year} {2018})}\BibitemShut {NoStop}%
\bibitem [{\citenamefont {Holstein}(1959)}]{Ho59a}%
  \BibitemOpen
  \bibfield  {author} {\bibinfo {author} {\bibfnamefont {T.}~\bibnamefont
  {Holstein}},\ }\href@noop {} {\bibfield  {journal} {\bibinfo  {journal} {Ann.
  Phys. (N.Y.)}\ }\textbf {\bibinfo {volume} {8}},\ \bibinfo {pages} {325
  (1959); {\bf 8}, 343} (\bibinfo {year} {1959})}\BibitemShut {NoStop}%
\bibitem [{\citenamefont {Blankenbecler}\ \emph {et~al.}(1981)\citenamefont
  {Blankenbecler}, \citenamefont {Scalapino},\ and\ \citenamefont
  {Sugar}}]{Blankenbecler81}%
  \BibitemOpen
  \bibfield  {author} {\bibinfo {author} {\bibfnamefont {R.}~\bibnamefont
  {Blankenbecler}}, \bibinfo {author} {\bibfnamefont {D.~J.}\ \bibnamefont
  {Scalapino}}, \ and\ \bibinfo {author} {\bibfnamefont {R.~L.}\ \bibnamefont
  {Sugar}},\ }\href@noop {} {\bibfield  {journal} {\bibinfo  {journal} {Phys.
  Rev. D}\ }\textbf {\bibinfo {volume} {24}},\ \bibinfo {pages} {2278}
  (\bibinfo {year} {1981})}\BibitemShut {NoStop}%
\bibitem [{\citenamefont {Rubtsov}\ \emph {et~al.}(2005)\citenamefont
  {Rubtsov}, \citenamefont {Savkin},\ and\ \citenamefont
  {Lichtenstein}}]{Rubtsov05}%
  \BibitemOpen
  \bibfield  {author} {\bibinfo {author} {\bibfnamefont {A.~N.}\ \bibnamefont
  {Rubtsov}}, \bibinfo {author} {\bibfnamefont {V.~V.}\ \bibnamefont {Savkin}},
  \ and\ \bibinfo {author} {\bibfnamefont {A.~I.}\ \bibnamefont
  {Lichtenstein}},\ }\href@noop {} {\bibfield  {journal} {\bibinfo  {journal}
  {Phys. Rev. B}\ }\textbf {\bibinfo {volume} {72}},\ \bibinfo {pages} {035122}
  (\bibinfo {year} {2005})}\BibitemShut {NoStop}%
\bibitem [{\citenamefont {Scalettar}\ \emph {et~al.}(1991)\citenamefont
  {Scalettar}, \citenamefont {Noack},\ and\ \citenamefont
  {Singh}}]{scalettar1991}%
  \BibitemOpen
  \bibfield  {author} {\bibinfo {author} {\bibfnamefont {R.~T.}\ \bibnamefont
  {Scalettar}}, \bibinfo {author} {\bibfnamefont {R.~M.}\ \bibnamefont
  {Noack}}, \ and\ \bibinfo {author} {\bibfnamefont {R.~R.~P.}\ \bibnamefont
  {Singh}},\ }\href {\doibase 10.1103/PhysRevB.44.10502} {\bibfield  {journal}
  {\bibinfo  {journal} {Phys. Rev. B}\ }\textbf {\bibinfo {volume} {44}},\
  \bibinfo {pages} {10502} (\bibinfo {year} {1991})}\BibitemShut {NoStop}%
\bibitem [{\citenamefont {Johnston}\ \emph {et~al.}(2013)\citenamefont
  {Johnston}, \citenamefont {Nowadnick}, \citenamefont {Kung}, \citenamefont
  {Moritz}, \citenamefont {Scalettar},\ and\ \citenamefont
  {Devereaux}}]{johnston2013}%
  \BibitemOpen
  \bibfield  {author} {\bibinfo {author} {\bibfnamefont {S.}~\bibnamefont
  {Johnston}}, \bibinfo {author} {\bibfnamefont {E.~A.}\ \bibnamefont
  {Nowadnick}}, \bibinfo {author} {\bibfnamefont {Y.~F.}\ \bibnamefont {Kung}},
  \bibinfo {author} {\bibfnamefont {B.}~\bibnamefont {Moritz}}, \bibinfo
  {author} {\bibfnamefont {R.~T.}\ \bibnamefont {Scalettar}}, \ and\ \bibinfo
  {author} {\bibfnamefont {T.~P.}\ \bibnamefont {Devereaux}},\ }\href {\doibase
  10.1103/PhysRevB.87.235133} {\bibfield  {journal} {\bibinfo  {journal} {Phys.
  Rev. B}\ }\textbf {\bibinfo {volume} {87}},\ \bibinfo {pages} {235133}
  (\bibinfo {year} {2013})}\BibitemShut {NoStop}%
\bibitem [{\citenamefont {Liu}\ \emph {et~al.}(2017{\natexlab{a}})\citenamefont
  {Liu}, \citenamefont {Qi}, \citenamefont {Meng},\ and\ \citenamefont
  {Fu}}]{PhysRevB.95.041101}%
  \BibitemOpen
  \bibfield  {author} {\bibinfo {author} {\bibfnamefont {J.}~\bibnamefont
  {Liu}}, \bibinfo {author} {\bibfnamefont {Y.}~\bibnamefont {Qi}}, \bibinfo
  {author} {\bibfnamefont {Z.~Y.}\ \bibnamefont {Meng}}, \ and\ \bibinfo
  {author} {\bibfnamefont {L.}~\bibnamefont {Fu}},\ }\href {\doibase
  10.1103/PhysRevB.95.041101} {\bibfield  {journal} {\bibinfo  {journal} {Phys.
  Rev. B}\ }\textbf {\bibinfo {volume} {95}},\ \bibinfo {pages} {041101}
  (\bibinfo {year} {2017}{\natexlab{a}})}\BibitemShut {NoStop}%
\bibitem [{\citenamefont {Liu}\ \emph {et~al.}(2017{\natexlab{b}})\citenamefont
  {Liu}, \citenamefont {Shen}, \citenamefont {Qi}, \citenamefont {Meng},\ and\
  \citenamefont {Fu}}]{PhysRevB.95.241104}%
  \BibitemOpen
  \bibfield  {author} {\bibinfo {author} {\bibfnamefont {J.}~\bibnamefont
  {Liu}}, \bibinfo {author} {\bibfnamefont {H.}~\bibnamefont {Shen}}, \bibinfo
  {author} {\bibfnamefont {Y.}~\bibnamefont {Qi}}, \bibinfo {author}
  {\bibfnamefont {Z.~Y.}\ \bibnamefont {Meng}}, \ and\ \bibinfo {author}
  {\bibfnamefont {L.}~\bibnamefont {Fu}},\ }\href {\doibase
  10.1103/PhysRevB.95.241104} {\bibfield  {journal} {\bibinfo  {journal} {Phys.
  Rev. B}\ }\textbf {\bibinfo {volume} {95}},\ \bibinfo {pages} {241104}
  (\bibinfo {year} {2017}{\natexlab{b}})}\BibitemShut {NoStop}%
\bibitem [{\citenamefont {Xu}\ \emph {et~al.}(2017)\citenamefont {Xu},
  \citenamefont {Qi}, \citenamefont {Liu}, \citenamefont {Fu},\ and\
  \citenamefont {Meng}}]{Xu2017SLMC}%
  \BibitemOpen
  \bibfield  {author} {\bibinfo {author} {\bibfnamefont {X.~Y.}\ \bibnamefont
  {Xu}}, \bibinfo {author} {\bibfnamefont {Y.}~\bibnamefont {Qi}}, \bibinfo
  {author} {\bibfnamefont {J.}~\bibnamefont {Liu}}, \bibinfo {author}
  {\bibfnamefont {L.}~\bibnamefont {Fu}}, \ and\ \bibinfo {author}
  {\bibfnamefont {Z.~Y.}\ \bibnamefont {Meng}},\ }\href {\doibase
  10.1103/PhysRevB.96.041119} {\bibfield  {journal} {\bibinfo  {journal} {Phys.
  Rev. B}\ }\textbf {\bibinfo {volume} {96}},\ \bibinfo {pages} {041119}
  (\bibinfo {year} {2017})}\BibitemShut {NoStop}%
\bibitem [{\citenamefont {Nagai}\ \emph {et~al.}(2017)\citenamefont {Nagai},
  \citenamefont {Shen}, \citenamefont {Qi}, \citenamefont {Liu},\ and\
  \citenamefont {Fu}}]{Nagaiself2017}%
  \BibitemOpen
  \bibfield  {author} {\bibinfo {author} {\bibfnamefont {Y.}~\bibnamefont
  {Nagai}}, \bibinfo {author} {\bibfnamefont {H.}~\bibnamefont {Shen}},
  \bibinfo {author} {\bibfnamefont {Y.}~\bibnamefont {Qi}}, \bibinfo {author}
  {\bibfnamefont {J.}~\bibnamefont {Liu}}, \ and\ \bibinfo {author}
  {\bibfnamefont {L.}~\bibnamefont {Fu}},\ }\href {\doibase
  10.1103/PhysRevB.96.161102} {\bibfield  {journal} {\bibinfo  {journal} {Phys.
  Rev. B}\ }\textbf {\bibinfo {volume} {96}},\ \bibinfo {pages} {161102}
  (\bibinfo {year} {2017})}\BibitemShut {NoStop}%
\bibitem [{SM()}]{SM}%
  \BibitemOpen
  \href@noop {} {}\bibinfo {note} {See Supplemental Material, which
  includes
  Refs.~\cite{PhysRevB.95.041101,PhysRevB.95.241104,Xu2017SLMC,PhysRevB.98.041102,Nagaiself2017,LiHuang2017a,LiHuang2017b,PhysRevB.98.045116,2018arXiv180100127L,2018arXiv180808878L,liu2016self,
  liu2016fermion,SwendsenWang1987,Wolff1989,duane1987hybrid,Beach04a,ALF17}.}\BibitemShut {Stop}%
\bibitem [{\citenamefont {Assaad}\ and\ \citenamefont {Lang}(2007)}]{Assaad07}%
  \BibitemOpen
  \bibfield  {author} {\bibinfo {author} {\bibfnamefont {F.~F.}\ \bibnamefont
  {Assaad}}\ and\ \bibinfo {author} {\bibfnamefont {T.~C.}\ \bibnamefont
  {Lang}},\ }\href {\doibase 10.1103/PhysRevB.76.035116} {\bibfield  {journal}
  {\bibinfo  {journal} {Phys. Rev. B}\ }\textbf {\bibinfo {volume} {76}},\
  \bibinfo {eid} {035116} (\bibinfo {year} {2007})}\BibitemShut {NoStop}%
\bibitem [{\citenamefont {Parisen~Toldin}\ \emph {et~al.}(2015)\citenamefont
  {Parisen~Toldin}, \citenamefont {Hohenadler}, \citenamefont {Assaad},\ and\
  \citenamefont {Herbut}}]{Toldin14}%
  \BibitemOpen
  \bibfield  {author} {\bibinfo {author} {\bibfnamefont {F.}~\bibnamefont
  {Parisen~Toldin}}, \bibinfo {author} {\bibfnamefont {M.}~\bibnamefont
  {Hohenadler}}, \bibinfo {author} {\bibfnamefont {F.~F.}\ \bibnamefont
  {Assaad}}, \ and\ \bibinfo {author} {\bibfnamefont {I.~F.}\ \bibnamefont
  {Herbut}},\ }\href {\doibase 10.1103/PhysRevB.91.165108} {\bibfield
  {journal} {\bibinfo  {journal} {Phys. Rev. B}\ }\textbf {\bibinfo {volume}
  {91}},\ \bibinfo {pages} {165108} (\bibinfo {year} {2015})}\BibitemShut
  {NoStop}%
\bibitem [{\citenamefont {Hirsch}\ and\ \citenamefont
  {Fradkin}(1983)}]{Hirsch83a}%
  \BibitemOpen
  \bibfield  {author} {\bibinfo {author} {\bibfnamefont {J.~E.}\ \bibnamefont
  {Hirsch}}\ and\ \bibinfo {author} {\bibfnamefont {E.}~\bibnamefont
  {Fradkin}},\ }\href@noop {} {\bibfield  {journal} {\bibinfo  {journal} {Phys.
  Rev. B}\ }\textbf {\bibinfo {volume} {27}},\ \bibinfo {pages} {4302}
  (\bibinfo {year} {1983})}\BibitemShut {NoStop}%
\bibitem [{\citenamefont {Yang}\ and\ \citenamefont
  {Zhang}(1990)}]{yang1990so}%
  \BibitemOpen
  \bibfield  {author} {\bibinfo {author} {\bibfnamefont {C.~N.}\ \bibnamefont
  {Yang}}\ and\ \bibinfo {author} {\bibfnamefont {S.}~\bibnamefont {Zhang}},\
  }\href@noop {} {\bibfield  {journal} {\bibinfo  {journal} {Mod. Phys. Lett.
  B}\ }\textbf {\bibinfo {volume} {04}},\ \bibinfo {pages} {759} (\bibinfo
  {year} {1990})}\BibitemShut {NoStop}%
\bibitem [{\citenamefont {Zhao}\ and\ \citenamefont
  {Paramekanti}(2006)}]{PhysRevLett.97.230404}%
  \BibitemOpen
  \bibfield  {author} {\bibinfo {author} {\bibfnamefont {E.}~\bibnamefont
  {Zhao}}\ and\ \bibinfo {author} {\bibfnamefont {A.}~\bibnamefont
  {Paramekanti}},\ }\href {\doibase 10.1103/PhysRevLett.97.230404} {\bibfield
  {journal} {\bibinfo  {journal} {Phys. Rev. Lett.}\ }\textbf {\bibinfo
  {volume} {97}},\ \bibinfo {pages} {230404} (\bibinfo {year}
  {2006})}\BibitemShut {NoStop}%
\bibitem [{\citenamefont {Hirsch}(1985)}]{Hirsch85}%
  \BibitemOpen
  \bibfield  {author} {\bibinfo {author} {\bibfnamefont {J.~E.}\ \bibnamefont
  {Hirsch}},\ }\href@noop {} {\bibfield  {journal} {\bibinfo  {journal} {Phys.
  Rev. B}\ }\textbf {\bibinfo {volume} {31}},\ \bibinfo {pages} {4403}
  (\bibinfo {year} {1985})}\BibitemShut {NoStop}%
\bibitem [{\citenamefont {Marsiglio}(1990)}]{PhysRevB.42.2416}%
  \BibitemOpen
  \bibfield  {author} {\bibinfo {author} {\bibfnamefont {F.}~\bibnamefont
  {Marsiglio}},\ }\href {\doibase 10.1103/PhysRevB.42.2416} {\bibfield
  {journal} {\bibinfo  {journal} {Phys. Rev. B}\ }\textbf {\bibinfo {volume}
  {42}},\ \bibinfo {pages} {2416} (\bibinfo {year} {1990})}\BibitemShut
  {NoStop}%
\bibitem [{\citenamefont {Binder}(1981)}]{Binder1981}%
  \BibitemOpen
  \bibfield  {author} {\bibinfo {author} {\bibfnamefont {K.}~\bibnamefont
  {Binder}},\ }\href {\doibase 10.1007/BF01293604} {\bibfield  {journal}
  {\bibinfo  {journal} {Z. Phys. B Con. Mat.}\ }\textbf {\bibinfo {volume}
  {43}},\ \bibinfo {pages} {119} (\bibinfo {year} {1981})}\BibitemShut
  {NoStop}%
\bibitem [{\citenamefont {Pujari}\ \emph {et~al.}(2016)\citenamefont {Pujari},
  \citenamefont {Lang}, \citenamefont {Murthy},\ and\ \citenamefont
  {Kaul}}]{PhysRevLett.117.086404}%
  \BibitemOpen
  \bibfield  {author} {\bibinfo {author} {\bibfnamefont {S.}~\bibnamefont
  {Pujari}}, \bibinfo {author} {\bibfnamefont {T.~C.}\ \bibnamefont {Lang}},
  \bibinfo {author} {\bibfnamefont {G.}~\bibnamefont {Murthy}}, \ and\ \bibinfo
  {author} {\bibfnamefont {R.~K.}\ \bibnamefont {Kaul}},\ }\href {\doibase
  10.1103/PhysRevLett.117.086404} {\bibfield  {journal} {\bibinfo  {journal}
  {Phys. Rev. Lett.}\ }\textbf {\bibinfo {volume} {117}},\ \bibinfo {pages}
  {086404} (\bibinfo {year} {2016})}\BibitemShut {NoStop}%
\bibitem [{\citenamefont {Herbut}\ \emph
  {et~al.}(2009{\natexlab{b}})\citenamefont {Herbut}, \citenamefont
  {Juri\ifmmode \check{c}\else \v{c}\fi{}i\ifmmode~\acute{c}\else \'{c}\fi{}},\
  and\ \citenamefont {Roy}}]{Herbut09}%
  \BibitemOpen
  \bibfield  {author} {\bibinfo {author} {\bibfnamefont {I.~F.}\ \bibnamefont
  {Herbut}}, \bibinfo {author} {\bibfnamefont {V.}~\bibnamefont {Juri\ifmmode
  \check{c}\else \v{c}\fi{}i\ifmmode~\acute{c}\else \'{c}\fi{}}}, \ and\
  \bibinfo {author} {\bibfnamefont {B.}~\bibnamefont {Roy}},\ }\href {\doibase
  10.1103/PhysRevB.79.085116} {\bibfield  {journal} {\bibinfo  {journal} {Phys.
  Rev. B}\ }\textbf {\bibinfo {volume} {79}},\ \bibinfo {pages} {085116}
  (\bibinfo {year} {2009}{\natexlab{b}})}\BibitemShut {NoStop}%
\bibitem [{\citenamefont {Blawid}\ and\ \citenamefont
  {Millis}(2001)}]{PhysRevB.63.115114}%
  \BibitemOpen
  \bibfield  {author} {\bibinfo {author} {\bibfnamefont {S.}~\bibnamefont
  {Blawid}}\ and\ \bibinfo {author} {\bibfnamefont {A.~J.}\ \bibnamefont
  {Millis}},\ }\href {\doibase 10.1103/PhysRevB.63.115114} {\bibfield
  {journal} {\bibinfo  {journal} {Phys. Rev. B}\ }\textbf {\bibinfo {volume}
  {63}},\ \bibinfo {pages} {115114} (\bibinfo {year} {2001})}\BibitemShut
  {NoStop}%
\bibitem [{\citenamefont {Gubernatis}\ \emph {et~al.}(1985)\citenamefont
  {Gubernatis}, \citenamefont {Scalapino}, \citenamefont {Sugar},\ and\
  \citenamefont {Toussaint}}]{PhysRevB.32.103}%
  \BibitemOpen
  \bibfield  {author} {\bibinfo {author} {\bibfnamefont {J.~E.}\ \bibnamefont
  {Gubernatis}}, \bibinfo {author} {\bibfnamefont {D.~J.}\ \bibnamefont
  {Scalapino}}, \bibinfo {author} {\bibfnamefont {R.~L.}\ \bibnamefont
  {Sugar}}, \ and\ \bibinfo {author} {\bibfnamefont {W.~D.}\ \bibnamefont
  {Toussaint}},\ }\href {\doibase 10.1103/PhysRevB.32.103} {\bibfield
  {journal} {\bibinfo  {journal} {Phys. Rev. B}\ }\textbf {\bibinfo {volume}
  {32}},\ \bibinfo {pages} {103} (\bibinfo {year} {1985})}\BibitemShut
  {NoStop}%
\bibitem [{\citenamefont {Bercx}\ \emph
  {et~al.}(2017{\natexlab{a}})\citenamefont {Bercx}, \citenamefont {Hofmann},
  \citenamefont {Assaad},\ and\ \citenamefont {Lang}}]{PhysRevB.95.035108}%
  \BibitemOpen
  \bibfield  {author} {\bibinfo {author} {\bibfnamefont {M.}~\bibnamefont
  {Bercx}}, \bibinfo {author} {\bibfnamefont {J.~S.}\ \bibnamefont {Hofmann}},
  \bibinfo {author} {\bibfnamefont {F.~F.}\ \bibnamefont {Assaad}}, \ and\
  \bibinfo {author} {\bibfnamefont {T.~C.}\ \bibnamefont {Lang}},\ }\href
  {\doibase 10.1103/PhysRevB.95.035108} {\bibfield  {journal} {\bibinfo
  {journal} {Phys. Rev. B}\ }\textbf {\bibinfo {volume} {95}},\ \bibinfo
  {pages} {035108} (\bibinfo {year} {2017}{\natexlab{a}})}\BibitemShut
  {NoStop}%
\bibitem [{\citenamefont {Otsuka}\ \emph {et~al.}(2016)\citenamefont {Otsuka},
  \citenamefont {Yunoki},\ and\ \citenamefont {Sorella}}]{Otsuka16}%
  \BibitemOpen
  \bibfield  {author} {\bibinfo {author} {\bibfnamefont {Y.}~\bibnamefont
  {Otsuka}}, \bibinfo {author} {\bibfnamefont {S.}~\bibnamefont {Yunoki}}, \
  and\ \bibinfo {author} {\bibfnamefont {S.}~\bibnamefont {Sorella}},\ }\href
  {\doibase 10.1103/PhysRevX.6.011029} {\bibfield  {journal} {\bibinfo
  {journal} {Phys. Rev. X}\ }\textbf {\bibinfo {volume} {6}},\ \bibinfo {pages}
  {011029} (\bibinfo {year} {2016})}\BibitemShut {NoStop}%
\bibitem [{\citenamefont {Zerf}\ \emph {et~al.}(2017)\citenamefont {Zerf},
  \citenamefont {Mihaila}, \citenamefont {Marquard}, \citenamefont {Herbut},\
  and\ \citenamefont {Scherer}}]{PhysRevD.96.096010}%
  \BibitemOpen
  \bibfield  {author} {\bibinfo {author} {\bibfnamefont {N.}~\bibnamefont
  {Zerf}}, \bibinfo {author} {\bibfnamefont {L.~N.}\ \bibnamefont {Mihaila}},
  \bibinfo {author} {\bibfnamefont {P.}~\bibnamefont {Marquard}}, \bibinfo
  {author} {\bibfnamefont {I.~F.}\ \bibnamefont {Herbut}}, \ and\ \bibinfo
  {author} {\bibfnamefont {M.~M.}\ \bibnamefont {Scherer}},\ }\href {\doibase
  10.1103/PhysRevD.96.096010} {\bibfield  {journal} {\bibinfo  {journal} {Phys.
  Rev. D}\ }\textbf {\bibinfo {volume} {96}},\ \bibinfo {pages} {096010}
  (\bibinfo {year} {2017})}\BibitemShut {NoStop}%
\bibitem [{\citenamefont {Melchert}(2009)}]{autoscale}%
  \BibitemOpen
  \bibfield  {author} {\bibinfo {author} {\bibfnamefont {O.}~\bibnamefont
  {Melchert}},\ }\href@noop {} {\bibfield  {journal} {\bibinfo  {journal}
  {arXiv:0910.5403}\ } (\bibinfo {year} {2009})}\BibitemShut {NoStop}%
\bibitem [{\citenamefont {Hohenadler}\ and\ \citenamefont
  {Assaad}(2013)}]{PhysRevB87.075149}%
  \BibitemOpen
  \bibfield  {author} {\bibinfo {author} {\bibfnamefont {M.}~\bibnamefont
  {Hohenadler}}\ and\ \bibinfo {author} {\bibfnamefont {F.~F.}\ \bibnamefont
  {Assaad}},\ }\href@noop {} {\bibfield  {journal} {\bibinfo  {journal} {Phys.
  Rev. B}\ }\textbf {\bibinfo {volume} {87}},\ \bibinfo {pages} {075149}
  (\bibinfo {year} {2013})}\BibitemShut {NoStop}%
\bibitem [{\citenamefont {Hasenbusch}\ \emph {et~al.}(1999)\citenamefont
  {Hasenbusch}, \citenamefont {Pinn},\ and\ \citenamefont
  {Vinti}}]{PhysRevB.59.11471}%
  \BibitemOpen
  \bibfield  {author} {\bibinfo {author} {\bibfnamefont {M.}~\bibnamefont
  {Hasenbusch}}, \bibinfo {author} {\bibfnamefont {K.}~\bibnamefont {Pinn}}, \
  and\ \bibinfo {author} {\bibfnamefont {S.}~\bibnamefont {Vinti}},\ }\href
  {\doibase 10.1103/PhysRevB.59.11471} {\bibfield  {journal} {\bibinfo
  {journal} {Phys. Rev. B}\ }\textbf {\bibinfo {volume} {59}},\ \bibinfo
  {pages} {11471} (\bibinfo {year} {1999})}\BibitemShut {NoStop}%
\bibitem [{\citenamefont {Costa}\ \emph {et~al.}(2018)\citenamefont {Costa},
  \citenamefont {Blommel}, \citenamefont {Chiu}, \citenamefont {Batrouni},\
  and\ \citenamefont {Scalettar}}]{PhysRevLett.120.187003}%
  \BibitemOpen
  \bibfield  {author} {\bibinfo {author} {\bibfnamefont {N.~C.}\ \bibnamefont
  {Costa}}, \bibinfo {author} {\bibfnamefont {T.}~\bibnamefont {Blommel}},
  \bibinfo {author} {\bibfnamefont {W.-T.}\ \bibnamefont {Chiu}}, \bibinfo
  {author} {\bibfnamefont {G.}~\bibnamefont {Batrouni}}, \ and\ \bibinfo
  {author} {\bibfnamefont {R.~T.}\ \bibnamefont {Scalettar}},\ }\href {\doibase
  10.1103/PhysRevLett.120.187003} {\bibfield  {journal} {\bibinfo  {journal}
  {Phys. Rev. Lett.}\ }\textbf {\bibinfo {volume} {120}},\ \bibinfo {pages}
  {187003} (\bibinfo {year} {2018})}\BibitemShut {NoStop}%
\bibitem [{\citenamefont {{J\"ulich Supercomputing Centre
  (2016)}}()}]{Juelich}%
  \BibitemOpen
  \bibfield  {author} {\bibinfo {author} {\bibnamefont {{J\"ulich
  Supercomputing Centre (2016)}}},\ }in\ \href@noop {} {\emph {\bibinfo
  {booktitle} {JURECA: General-purpose supercomputer at J\"ulich Supercomputing
  Centre}}},\ \bibinfo {series} {Journal of Large-Scale Research Facilities},
  Vol.~\bibinfo {volume} {2},\ p.\ \bibinfo {pages} {A62},\ \bibinfo {note}
  {http://dx.doi.org/10.17815/jlsrf-2-121}\BibitemShut {NoStop}%
\bibitem [{\citenamefont {Huang}\ and\ \citenamefont
  {Wang}(2017)}]{LiHuang2017a}%
  \BibitemOpen
  \bibfield  {author} {\bibinfo {author} {\bibfnamefont {L.}~\bibnamefont
  {Huang}}\ and\ \bibinfo {author} {\bibfnamefont {L.}~\bibnamefont {Wang}},\
  }\href {\doibase 10.1103/PhysRevB.95.035105} {\bibfield  {journal} {\bibinfo
  {journal} {Phys. Rev. B}\ }\textbf {\bibinfo {volume} {95}},\ \bibinfo
  {pages} {035105} (\bibinfo {year} {2017})}\BibitemShut {NoStop}%
\bibitem [{\citenamefont {Huang}\ \emph {et~al.}(2017)\citenamefont {Huang},
  \citenamefont {Yang},\ and\ \citenamefont {Wang}}]{LiHuang2017b}%
  \BibitemOpen
  \bibfield  {author} {\bibinfo {author} {\bibfnamefont {L.}~\bibnamefont
  {Huang}}, \bibinfo {author} {\bibfnamefont {Y.-F.}\ \bibnamefont {Yang}}, \
  and\ \bibinfo {author} {\bibfnamefont {L.}~\bibnamefont {Wang}},\ }\href
  {\doibase 10.1103/PhysRevE.95.031301} {\bibfield  {journal} {\bibinfo
  {journal} {Phys. Rev. E}\ }\textbf {\bibinfo {volume} {95}},\ \bibinfo
  {pages} {031301} (\bibinfo {year} {2017})}\BibitemShut {NoStop}%
\bibitem [{\citenamefont {Liu}\ \emph {et~al.}(2018)\citenamefont {Liu},
  \citenamefont {Xu}, \citenamefont {Qi}, \citenamefont {Sun},\ and\
  \citenamefont {Meng}}]{PhysRevB.98.045116}%
  \BibitemOpen
  \bibfield  {author} {\bibinfo {author} {\bibfnamefont {Z.~H.}\ \bibnamefont
  {Liu}}, \bibinfo {author} {\bibfnamefont {X.~Y.}\ \bibnamefont {Xu}},
  \bibinfo {author} {\bibfnamefont {Y.}~\bibnamefont {Qi}}, \bibinfo {author}
  {\bibfnamefont {K.}~\bibnamefont {Sun}}, \ and\ \bibinfo {author}
  {\bibfnamefont {Z.~Y.}\ \bibnamefont {Meng}},\ }\href {\doibase
  10.1103/PhysRevB.98.045116} {\bibfield  {journal} {\bibinfo  {journal} {Phys.
  Rev. B}\ }\textbf {\bibinfo {volume} {98}},\ \bibinfo {pages} {045116}
  (\bibinfo {year} {2018})}\BibitemShut {NoStop}%
\bibitem [{\citenamefont {{Liu}}\ \emph
  {et~al.}(2018{\natexlab{a}})\citenamefont {{Liu}}, \citenamefont {{Xu}},
  \citenamefont {{Qi}}, \citenamefont {{Sun}},\ and\ \citenamefont
  {{Meng}}}]{2018arXiv180100127L}%
  \BibitemOpen
  \bibfield  {author} {\bibinfo {author} {\bibfnamefont {Z.~H.}\ \bibnamefont
  {{Liu}}}, \bibinfo {author} {\bibfnamefont {X.~Y.}\ \bibnamefont {{Xu}}},
  \bibinfo {author} {\bibfnamefont {Y.}~\bibnamefont {{Qi}}}, \bibinfo {author}
  {\bibfnamefont {K.}~\bibnamefont {{Sun}}}, \ and\ \bibinfo {author}
  {\bibfnamefont {Z.~Y.}\ \bibnamefont {{Meng}}},\ }\href@noop {} {\bibfield
  {journal} {\bibinfo  {journal} {ArXiv e-prints}\ } (\bibinfo {year}
  {2018}{\natexlab{a}})},\ \Eprint {http://arxiv.org/abs/1801.00127}
  {arXiv:1801.00127} \BibitemShut {NoStop}%
\bibitem [{\citenamefont {{Liu}}\ \emph
  {et~al.}(2018{\natexlab{b}})\citenamefont {{Liu}}, \citenamefont {{Pan}},
  \citenamefont {{Xu}}, \citenamefont {{Sun}},\ and\ \citenamefont
  {{Meng}}}]{2018arXiv180808878L}%
  \BibitemOpen
  \bibfield  {author} {\bibinfo {author} {\bibfnamefont {Z.~H.}\ \bibnamefont
  {{Liu}}}, \bibinfo {author} {\bibfnamefont {G.}~\bibnamefont {{Pan}}},
  \bibinfo {author} {\bibfnamefont {X.~Y.}\ \bibnamefont {{Xu}}}, \bibinfo
  {author} {\bibfnamefont {K.}~\bibnamefont {{Sun}}}, \ and\ \bibinfo {author}
  {\bibfnamefont {Z.~Y.}\ \bibnamefont {{Meng}}},\ }\href@noop {} {\bibfield
  {journal} {\bibinfo  {journal} {ArXiv e-prints}\ } (\bibinfo {year}
  {2018}{\natexlab{b}})},\ \Eprint {http://arxiv.org/abs/1808.08878}
  {arXiv:1808.08878} \BibitemShut {NoStop}%
\bibitem [{\citenamefont {Liu}\ \emph {et~al.}(2017{\natexlab{c}})\citenamefont
  {Liu}, \citenamefont {Qi}, \citenamefont {Meng},\ and\ \citenamefont
  {Fu}}]{liu2016self}%
  \BibitemOpen
  \bibfield  {author} {\bibinfo {author} {\bibfnamefont {J.}~\bibnamefont
  {Liu}}, \bibinfo {author} {\bibfnamefont {Y.}~\bibnamefont {Qi}}, \bibinfo
  {author} {\bibfnamefont {Z.~Y.}\ \bibnamefont {Meng}}, \ and\ \bibinfo
  {author} {\bibfnamefont {L.}~\bibnamefont {Fu}},\ }\href {\doibase
  10.1103/PhysRevB.95.041101} {\bibfield  {journal} {\bibinfo  {journal} {Phys.
  Rev. B}\ }\textbf {\bibinfo {volume} {95}},\ \bibinfo {pages} {041101}
  (\bibinfo {year} {2017}{\natexlab{c}})}\BibitemShut {NoStop}%
\bibitem [{\citenamefont {Liu}\ \emph {et~al.}(2017{\natexlab{d}})\citenamefont
  {Liu}, \citenamefont {Shen}, \citenamefont {Qi}, \citenamefont {Meng},\ and\
  \citenamefont {Fu}}]{liu2016fermion}%
  \BibitemOpen
  \bibfield  {author} {\bibinfo {author} {\bibfnamefont {J.}~\bibnamefont
  {Liu}}, \bibinfo {author} {\bibfnamefont {H.}~\bibnamefont {Shen}}, \bibinfo
  {author} {\bibfnamefont {Y.}~\bibnamefont {Qi}}, \bibinfo {author}
  {\bibfnamefont {Z.~Y.}\ \bibnamefont {Meng}}, \ and\ \bibinfo {author}
  {\bibfnamefont {L.}~\bibnamefont {Fu}},\ }\href {\doibase
  10.1103/PhysRevB.95.241104} {\bibfield  {journal} {\bibinfo  {journal} {Phys.
  Rev. B}\ }\textbf {\bibinfo {volume} {95}},\ \bibinfo {pages} {241104}
  (\bibinfo {year} {2017}{\natexlab{d}})}\BibitemShut {NoStop}%
\bibitem [{\citenamefont {Swendsen}\ and\ \citenamefont
  {Wang}(1987)}]{SwendsenWang1987}%
  \BibitemOpen
  \bibfield  {author} {\bibinfo {author} {\bibfnamefont {R.~H.}\ \bibnamefont
  {Swendsen}}\ and\ \bibinfo {author} {\bibfnamefont {J.-S.}\ \bibnamefont
  {Wang}},\ }\href {\doibase 10.1103/PhysRevLett.58.86} {\bibfield  {journal}
  {\bibinfo  {journal} {Phys. Rev. Lett.}\ }\textbf {\bibinfo {volume} {58}},\
  \bibinfo {pages} {86} (\bibinfo {year} {1987})}\BibitemShut {NoStop}%
\bibitem [{\citenamefont {Wolff}(1989)}]{Wolff1989}%
  \BibitemOpen
  \bibfield  {author} {\bibinfo {author} {\bibfnamefont {U.}~\bibnamefont
  {Wolff}},\ }\href {\doibase 10.1103/PhysRevLett.62.361} {\bibfield  {journal}
  {\bibinfo  {journal} {Phys. Rev. Lett.}\ }\textbf {\bibinfo {volume} {62}},\
  \bibinfo {pages} {361} (\bibinfo {year} {1989})}\BibitemShut {NoStop}%
\bibitem [{\citenamefont {Duane}\ \emph {et~al.}(1987)\citenamefont {Duane},
  \citenamefont {Kennedy}, \citenamefont {Pendleton},\ and\ \citenamefont
  {Roweth}}]{duane1987hybrid}%
  \BibitemOpen
  \bibfield  {author} {\bibinfo {author} {\bibfnamefont {S.}~\bibnamefont
  {Duane}}, \bibinfo {author} {\bibfnamefont {A.~D.}\ \bibnamefont {Kennedy}},
  \bibinfo {author} {\bibfnamefont {B.~J.}\ \bibnamefont {Pendleton}}, \ and\
  \bibinfo {author} {\bibfnamefont {D.}~\bibnamefont {Roweth}},\ }\href@noop {}
  {\bibfield  {journal} {\bibinfo  {journal} {Phys. Lett. B}\ }\textbf
  {\bibinfo {volume} {195}},\ \bibinfo {pages} {216} (\bibinfo {year}
  {1987})}\BibitemShut {NoStop}%
\bibitem [{\citenamefont {Beach}(2004)}]{Beach04a}%
  \BibitemOpen
  \bibfield  {author} {\bibinfo {author} {\bibfnamefont {K.~S.~D.}\
  \bibnamefont {Beach}},\ }\href@noop {} {\bibfield  {journal} {\bibinfo
  {journal} {arXiv:cond-mat/0403055}\ } (\bibinfo {year} {2004})}\BibitemShut
  {NoStop}%
\bibitem [{\citenamefont {Bercx}\ \emph
  {et~al.}(2017{\natexlab{b}})\citenamefont {Bercx}, \citenamefont {Goth},
  \citenamefont {Hofmann},\ and\ \citenamefont {Assaad}}]{ALF17}%
  \BibitemOpen
  \bibfield  {author} {\bibinfo {author} {\bibfnamefont {M.}~\bibnamefont
  {Bercx}}, \bibinfo {author} {\bibfnamefont {F.}~\bibnamefont {Goth}},
  \bibinfo {author} {\bibfnamefont {J.~S.}\ \bibnamefont {Hofmann}}, \ and\
  \bibinfo {author} {\bibfnamefont {F.~F.}\ \bibnamefont {Assaad}},\
  }\href@noop {} {\bibfield  {journal} {\bibinfo  {journal} {SciPost}\ }\textbf
  {\bibinfo {volume} {3}},\ \bibinfo {pages} {013} (\bibinfo {year}
  {2017}{\natexlab{b}})}\BibitemShut {NoStop}%
\bibitem [{\citenamefont {Zhang}\ \emph {et~al.}(2019)\citenamefont {Zhang},
  \citenamefont {Chiu}, \citenamefont {Costa}, \citenamefont {Batrouni},\ and\
  \citenamefont {Scalettar}}]{PhysRevLett.122.077602}%
  \BibitemOpen
  \bibfield  {author} {\bibinfo {author} {\bibfnamefont {Y.-X.}\ \bibnamefont
  {Zhang}}, \bibinfo {author} {\bibfnamefont {W.-T.}\ \bibnamefont {Chiu}},
  \bibinfo {author} {\bibfnamefont {N.~C.}\ \bibnamefont {Costa}}, \bibinfo
  {author} {\bibfnamefont {G.~G.}\ \bibnamefont {Batrouni}}, \ and\ \bibinfo
  {author} {\bibfnamefont {R.~T.}\ \bibnamefont {Scalettar}},\ }\href {\doibase
  10.1103/PhysRevLett.122.077602} {\bibfield  {journal} {\bibinfo  {journal}
  {Phys. Rev. Lett.}\ }\textbf {\bibinfo {volume} {122}},\ \bibinfo {pages}
  {077602} (\bibinfo {year} {2019})}\BibitemShut {NoStop}%

\end{thebibliography}

\begin{thebibliography}{17}%
\makeatletter
\providecommand \@ifxundefined [1]{%
 \@ifx{#1\undefined}
}%
\providecommand \@ifnum [1]{%
 \ifnum #1\expandafter \@firstoftwo
 \else \expandafter \@secondoftwo
 \fi
}%
\providecommand \@ifx [1]{%
 \ifx #1\expandafter \@firstoftwo
 \else \expandafter \@secondoftwo
 \fi
}%
\providecommand \natexlab [1]{#1}%
\providecommand \enquote  [1]{``#1''}%
\providecommand \bibnamefont  [1]{#1}%
\providecommand \bibfnamefont [1]{#1}%
\providecommand \citenamefont [1]{#1}%
\providecommand \href@noop [0]{\@secondoftwo}%
\providecommand \href [0]{\begingroup \@sanitize@url \@href}%
\providecommand \@href[1]{\@@startlink{#1}\@@href}%
\providecommand \@@href[1]{\endgroup#1\@@endlink}%
\providecommand \@sanitize@url [0]{\catcode `\\12\catcode `\$12\catcode
  `\&12\catcode `\#12\catcode `\^12\catcode `\_12\catcode `\%12\relax}%
\providecommand \@@startlink[1]{}%
\providecommand \@@endlink[0]{}%
\providecommand \url  [0]{\begingroup\@sanitize@url \@url }%
\providecommand \@url [1]{\endgroup\@href {#1}{\urlprefix }}%
\providecommand \urlprefix  [0]{URL }%
\providecommand \Eprint [0]{\href }%
\providecommand \doibase [0]{http://dx.doi.org/}%
\providecommand \selectlanguage [0]{\@gobble}%
\providecommand \bibinfo  [0]{\@secondoftwo}%
\providecommand \bibfield  [0]{\@secondoftwo}%
\providecommand \translation [1]{[#1]}%
\providecommand \BibitemOpen [0]{}%
\providecommand \bibitemStop [0]{}%
\providecommand \bibitemNoStop [0]{.\EOS\space}%
\providecommand \EOS [0]{\spacefactor3000\relax}%
\providecommand \BibitemShut  [1]{\csname bibitem#1\endcsname}%
\let\auto@bib@innerbib\@empty
\bibitem [{\citenamefont {Liu}\ \emph {et~al.}(2017{\natexlab{a}})\citenamefont
  {Liu}, \citenamefont {Qi}, \citenamefont {Meng},\ and\ \citenamefont
  {Fu}}]{PhysRevB.95.041101}%
  \BibitemOpen
  \bibfield  {author} {\bibinfo {author} {\bibfnamefont {J.}~\bibnamefont
  {Liu}}, \bibinfo {author} {\bibfnamefont {Y.}~\bibnamefont {Qi}}, \bibinfo
  {author} {\bibfnamefont {Z.~Y.}\ \bibnamefont {Meng}}, \ and\ \bibinfo
  {author} {\bibfnamefont {L.}~\bibnamefont {Fu}},\ }\href {\doibase
  10.1103/PhysRevB.95.041101} {\bibfield  {journal} {\bibinfo  {journal} {Phys.
  Rev. B}\ }\textbf {\bibinfo {volume} {95}},\ \bibinfo {pages} {041101}
  (\bibinfo {year} {2017}{\natexlab{a}})}\BibitemShut {NoStop}%
\bibitem [{\citenamefont {Liu}\ \emph {et~al.}(2017{\natexlab{b}})\citenamefont
  {Liu}, \citenamefont {Shen}, \citenamefont {Qi}, \citenamefont {Meng},\ and\
  \citenamefont {Fu}}]{PhysRevB.95.241104}%
  \BibitemOpen
  \bibfield  {author} {\bibinfo {author} {\bibfnamefont {J.}~\bibnamefont
  {Liu}}, \bibinfo {author} {\bibfnamefont {H.}~\bibnamefont {Shen}}, \bibinfo
  {author} {\bibfnamefont {Y.}~\bibnamefont {Qi}}, \bibinfo {author}
  {\bibfnamefont {Z.~Y.}\ \bibnamefont {Meng}}, \ and\ \bibinfo {author}
  {\bibfnamefont {L.}~\bibnamefont {Fu}},\ }\href {\doibase
  10.1103/PhysRevB.95.241104} {\bibfield  {journal} {\bibinfo  {journal} {Phys.
  Rev. B}\ }\textbf {\bibinfo {volume} {95}},\ \bibinfo {pages} {241104}
  (\bibinfo {year} {2017}{\natexlab{b}})}\BibitemShut {NoStop}%
\bibitem [{\citenamefont {Xu}\ \emph {et~al.}(2017)\citenamefont {Xu},
  \citenamefont {Qi}, \citenamefont {Liu}, \citenamefont {Fu},\ and\
  \citenamefont {Meng}}]{Xu2017SLMC}%
  \BibitemOpen
  \bibfield  {author} {\bibinfo {author} {\bibfnamefont {X.~Y.}\ \bibnamefont
  {Xu}}, \bibinfo {author} {\bibfnamefont {Y.}~\bibnamefont {Qi}}, \bibinfo
  {author} {\bibfnamefont {J.}~\bibnamefont {Liu}}, \bibinfo {author}
  {\bibfnamefont {L.}~\bibnamefont {Fu}}, \ and\ \bibinfo {author}
  {\bibfnamefont {Z.~Y.}\ \bibnamefont {Meng}},\ }\href {\doibase
  10.1103/PhysRevB.96.041119} {\bibfield  {journal} {\bibinfo  {journal} {Phys.
  Rev. B}\ }\textbf {\bibinfo {volume} {96}},\ \bibinfo {pages} {041119}
  (\bibinfo {year} {2017})}\BibitemShut {NoStop}%
\bibitem [{\citenamefont {Chen}\ \emph {et~al.}(2018)\citenamefont {Chen},
  \citenamefont {Xu}, \citenamefont {Liu}, \citenamefont {Batrouni},
  \citenamefont {Scalettar},\ and\ \citenamefont {Meng}}]{PhysRevB.98.041102}%
  \BibitemOpen
  \bibfield  {author} {\bibinfo {author} {\bibfnamefont {C.}~\bibnamefont
  {Chen}}, \bibinfo {author} {\bibfnamefont {X.~Y.}\ \bibnamefont {Xu}},
  \bibinfo {author} {\bibfnamefont {J.}~\bibnamefont {Liu}}, \bibinfo {author}
  {\bibfnamefont {G.}~\bibnamefont {Batrouni}}, \bibinfo {author}
  {\bibfnamefont {R.}~\bibnamefont {Scalettar}}, \ and\ \bibinfo {author}
  {\bibfnamefont {Z.~Y.}\ \bibnamefont {Meng}},\ }\href {\doibase
  10.1103/PhysRevB.98.041102} {\bibfield  {journal} {\bibinfo  {journal} {Phys.
  Rev. B}\ }\textbf {\bibinfo {volume} {98}},\ \bibinfo {pages} {041102}
  (\bibinfo {year} {2018})}\BibitemShut {NoStop}%
\bibitem [{\citenamefont {Nagai}\ \emph {et~al.}(2017)\citenamefont {Nagai},
  \citenamefont {Shen}, \citenamefont {Qi}, \citenamefont {Liu},\ and\
  \citenamefont {Fu}}]{Nagaiself2017}%
  \BibitemOpen
  \bibfield  {author} {\bibinfo {author} {\bibfnamefont {Y.}~\bibnamefont
  {Nagai}}, \bibinfo {author} {\bibfnamefont {H.}~\bibnamefont {Shen}},
  \bibinfo {author} {\bibfnamefont {Y.}~\bibnamefont {Qi}}, \bibinfo {author}
  {\bibfnamefont {J.}~\bibnamefont {Liu}}, \ and\ \bibinfo {author}
  {\bibfnamefont {L.}~\bibnamefont {Fu}},\ }\href {\doibase
  10.1103/PhysRevB.96.161102} {\bibfield  {journal} {\bibinfo  {journal} {Phys.
  Rev. B}\ }\textbf {\bibinfo {volume} {96}},\ \bibinfo {pages} {161102}
  (\bibinfo {year} {2017})}\BibitemShut {NoStop}%
\bibitem [{\citenamefont {Huang}\ and\ \citenamefont
  {Wang}(2017)}]{LiHuang2017a}%
  \BibitemOpen
  \bibfield  {author} {\bibinfo {author} {\bibfnamefont {L.}~\bibnamefont
  {Huang}}\ and\ \bibinfo {author} {\bibfnamefont {L.}~\bibnamefont {Wang}},\
  }\href {\doibase 10.1103/PhysRevB.95.035105} {\bibfield  {journal} {\bibinfo
  {journal} {Phys. Rev. B}\ }\textbf {\bibinfo {volume} {95}},\ \bibinfo
  {pages} {035105} (\bibinfo {year} {2017})}\BibitemShut {NoStop}%
\bibitem [{\citenamefont {Huang}\ \emph {et~al.}(2017)\citenamefont {Huang},
  \citenamefont {Yang},\ and\ \citenamefont {Wang}}]{LiHuang2017b}%
  \BibitemOpen
  \bibfield  {author} {\bibinfo {author} {\bibfnamefont {L.}~\bibnamefont
  {Huang}}, \bibinfo {author} {\bibfnamefont {Y.-f.}\ \bibnamefont {Yang}}, \
  and\ \bibinfo {author} {\bibfnamefont {L.}~\bibnamefont {Wang}},\ }\href
  {\doibase 10.1103/PhysRevE.95.031301} {\bibfield  {journal} {\bibinfo
  {journal} {Phys. Rev. E}\ }\textbf {\bibinfo {volume} {95}},\ \bibinfo
  {pages} {031301} (\bibinfo {year} {2017})}\BibitemShut {NoStop}%
\bibitem [{\citenamefont {Liu}\ \emph {et~al.}(2018)\citenamefont {Liu},
  \citenamefont {Xu}, \citenamefont {Qi}, \citenamefont {Sun},\ and\
  \citenamefont {Meng}}]{PhysRevB.98.045116}%
  \BibitemOpen
  \bibfield  {author} {\bibinfo {author} {\bibfnamefont {Z.~H.}\ \bibnamefont
  {Liu}}, \bibinfo {author} {\bibfnamefont {X.~Y.}\ \bibnamefont {Xu}},
  \bibinfo {author} {\bibfnamefont {Y.}~\bibnamefont {Qi}}, \bibinfo {author}
  {\bibfnamefont {K.}~\bibnamefont {Sun}}, \ and\ \bibinfo {author}
  {\bibfnamefont {Z.~Y.}\ \bibnamefont {Meng}},\ }\href {\doibase
  10.1103/PhysRevB.98.045116} {\bibfield  {journal} {\bibinfo  {journal} {Phys.
  Rev. B}\ }\textbf {\bibinfo {volume} {98}},\ \bibinfo {pages} {045116}
  (\bibinfo {year} {2018})}\BibitemShut {NoStop}%
\bibitem [{\citenamefont {{Liu}}\ \emph
  {et~al.}(2018{\natexlab{a}})\citenamefont {{Liu}}, \citenamefont {{Xu}},
  \citenamefont {{Qi}}, \citenamefont {{Sun}},\ and\ \citenamefont
  {{Meng}}}]{2018arXiv180100127L}%
  \BibitemOpen
  \bibfield  {author} {\bibinfo {author} {\bibfnamefont {Z.~H.}\ \bibnamefont
  {{Liu}}}, \bibinfo {author} {\bibfnamefont {X.~Y.}\ \bibnamefont {{Xu}}},
  \bibinfo {author} {\bibfnamefont {Y.}~\bibnamefont {{Qi}}}, \bibinfo {author}
  {\bibfnamefont {K.}~\bibnamefont {{Sun}}}, \ and\ \bibinfo {author}
  {\bibfnamefont {Z.~Y.}\ \bibnamefont {{Meng}}},\ }\href@noop {} {\bibfield
  {journal} {\bibinfo  {journal} {ArXiv e-prints}\ } (\bibinfo {year}
  {2018}{\natexlab{a}})},\ \Eprint {http://arxiv.org/abs/1801.00127}
  {arXiv:1801.00127} \BibitemShut {NoStop}%
\bibitem [{\citenamefont {{Liu}}\ \emph
  {et~al.}(2018{\natexlab{b}})\citenamefont {{Liu}}, \citenamefont {{Pan}},
  \citenamefont {{Xu}}, \citenamefont {{Sun}},\ and\ \citenamefont
  {{Meng}}}]{2018arXiv180808878L}%
  \BibitemOpen
  \bibfield  {author} {\bibinfo {author} {\bibfnamefont {Z.~H.}\ \bibnamefont
  {{Liu}}}, \bibinfo {author} {\bibfnamefont {G.}~\bibnamefont {{Pan}}},
  \bibinfo {author} {\bibfnamefont {X.~Y.}\ \bibnamefont {{Xu}}}, \bibinfo
  {author} {\bibfnamefont {K.}~\bibnamefont {{Sun}}}, \ and\ \bibinfo {author}
  {\bibfnamefont {Z.~Y.}\ \bibnamefont {{Meng}}},\ }\href@noop {} {\bibfield
  {journal} {\bibinfo  {journal} {ArXiv e-prints}\ } (\bibinfo {year}
  {2018}{\natexlab{b}})},\ \Eprint {http://arxiv.org/abs/1808.08878}
  {arXiv:1808.08878} \BibitemShut {NoStop}%
\bibitem [{\citenamefont {Liu}\ \emph {et~al.}(2017{\natexlab{c}})\citenamefont
  {Liu}, \citenamefont {Qi}, \citenamefont {Meng},\ and\ \citenamefont
  {Fu}}]{liu2016self}%
  \BibitemOpen
  \bibfield  {author} {\bibinfo {author} {\bibfnamefont {J.}~\bibnamefont
  {Liu}}, \bibinfo {author} {\bibfnamefont {Y.}~\bibnamefont {Qi}}, \bibinfo
  {author} {\bibfnamefont {Z.~Y.}\ \bibnamefont {Meng}}, \ and\ \bibinfo
  {author} {\bibfnamefont {L.}~\bibnamefont {Fu}},\ }\href {\doibase
  10.1103/PhysRevB.95.041101} {\bibfield  {journal} {\bibinfo  {journal} {Phys.
  Rev. B}\ }\textbf {\bibinfo {volume} {95}},\ \bibinfo {pages} {041101}
  (\bibinfo {year} {2017}{\natexlab{c}})}\BibitemShut {NoStop}%
\bibitem [{\citenamefont {Liu}\ \emph {et~al.}(2017{\natexlab{d}})\citenamefont
  {Liu}, \citenamefont {Shen}, \citenamefont {Qi}, \citenamefont {Meng},\ and\
  \citenamefont {Fu}}]{liu2016fermion}%
  \BibitemOpen
  \bibfield  {author} {\bibinfo {author} {\bibfnamefont {J.}~\bibnamefont
  {Liu}}, \bibinfo {author} {\bibfnamefont {H.}~\bibnamefont {Shen}}, \bibinfo
  {author} {\bibfnamefont {Y.}~\bibnamefont {Qi}}, \bibinfo {author}
  {\bibfnamefont {Z.~Y.}\ \bibnamefont {Meng}}, \ and\ \bibinfo {author}
  {\bibfnamefont {L.}~\bibnamefont {Fu}},\ }\href {\doibase
  10.1103/PhysRevB.95.241104} {\bibfield  {journal} {\bibinfo  {journal} {Phys.
  Rev. B}\ }\textbf {\bibinfo {volume} {95}},\ \bibinfo {pages} {241104}
  (\bibinfo {year} {2017}{\natexlab{d}})}\BibitemShut {NoStop}%
\bibitem [{\citenamefont {Swendsen}\ and\ \citenamefont
  {Wang}(1987)}]{SwendsenWang1987}%
  \BibitemOpen
  \bibfield  {author} {\bibinfo {author} {\bibfnamefont {R.~H.}\ \bibnamefont
  {Swendsen}}\ and\ \bibinfo {author} {\bibfnamefont {J.-S.}\ \bibnamefont
  {Wang}},\ }\href {\doibase 10.1103/PhysRevLett.58.86} {\bibfield  {journal}
  {\bibinfo  {journal} {Phys. Rev. Lett.}\ }\textbf {\bibinfo {volume} {58}},\
  \bibinfo {pages} {86} (\bibinfo {year} {1987})}\BibitemShut {NoStop}%
\bibitem [{\citenamefont {Wolff}(1989)}]{Wolff1989}%
  \BibitemOpen
  \bibfield  {author} {\bibinfo {author} {\bibfnamefont {U.}~\bibnamefont
  {Wolff}},\ }\href {\doibase 10.1103/PhysRevLett.62.361} {\bibfield  {journal}
  {\bibinfo  {journal} {Phys. Rev. Lett.}\ }\textbf {\bibinfo {volume} {62}},\
  \bibinfo {pages} {361} (\bibinfo {year} {1989})}\BibitemShut {NoStop}%
\bibitem [{\citenamefont {Duane}\ \emph {et~al.}(1987)\citenamefont {Duane},
  \citenamefont {Kennedy}, \citenamefont {Pendleton},\ and\ \citenamefont
  {Roweth}}]{duane1987hybrid}%
  \BibitemOpen
  \bibfield  {author} {\bibinfo {author} {\bibfnamefont {S.}~\bibnamefont
  {Duane}}, \bibinfo {author} {\bibfnamefont {A.~D.}\ \bibnamefont {Kennedy}},
  \bibinfo {author} {\bibfnamefont {B.~J.}\ \bibnamefont {Pendleton}}, \ and\
  \bibinfo {author} {\bibfnamefont {D.}~\bibnamefont {Roweth}},\ }\href@noop {}
  {\bibfield  {journal} {\bibinfo  {journal} {Physics Letters B}\ }\textbf
  {\bibinfo {volume} {195}},\ \bibinfo {pages} {216} (\bibinfo {year}
  {1987})}\BibitemShut {NoStop}%
\bibitem [{\citenamefont {Bercx}\ \emph {et~al.}(2017)\citenamefont {Bercx},
  \citenamefont {Goth}, \citenamefont {Hofmann},\ and\ \citenamefont
  {Assaad}}]{ALF17}%
  \BibitemOpen
  \bibfield  {author} {\bibinfo {author} {\bibfnamefont {M.}~\bibnamefont
  {Bercx}}, \bibinfo {author} {\bibfnamefont {F.}~\bibnamefont {Goth}},
  \bibinfo {author} {\bibfnamefont {J.~S.}\ \bibnamefont {Hofmann}}, \ and\
  \bibinfo {author} {\bibfnamefont {F.~F.}\ \bibnamefont {Assaad}},\
  }\href@noop {} {\bibfield  {journal} {\bibinfo  {journal} {SciPost}\ }\textbf
  {\bibinfo {volume} {3}},\ \bibinfo {pages} {013} (\bibinfo {year}
  {2017})}\BibitemShut {NoStop}%
\bibitem [{\citenamefont {Beach}(2004)}]{Beach04a}%
  \BibitemOpen
  \bibfield  {author} {\bibinfo {author} {\bibfnamefont {K.~S.~D.}\
  \bibnamefont {Beach}},\ }\href@noop {} {\bibfield  {journal} {\bibinfo
  {journal} {arXiv:cond-mat/0403055}\ } (\bibinfo {year} {2004})}\BibitemShut
  {NoStop}%
\end{thebibliography}

%

\end{document}